\documentclass[a4paper,fleqn,usenatbib,useAMS]{mnras}
\usepackage{graphicx}	% Including figure files
\usepackage{amsmath}	% Advanced maths commands
\usepackage{amssymb}	% Extra maths symbols

\newcommand{\Teff}{\ensuremath{T_{\rm eff}}}                % Effective temperature symbol
\newcommand{\logg}{\ensuremath{\log g}}                     % log(g) symbol
\newcommand{\Msun}{\ensuremath{\,{\rm M}_\odot}}            % Solar mass symbol
\newcommand{\Rsun}{\ensuremath{\,{\rm R}_\odot}}            % Solar radius symbol
\newcommand{\kms}{\,km\,s$^{-1}$}                           % km/s symbol
\newcommand{\spd}{{\sc spd}}

\title[Physical properties and CNO abundances in OB binaries]{Physical properties and CNO abundances for high-mass stars in four main-sequence detached eclipsing binaries: V478\,Cyg, AH\,Cep, V453\,Cyg and V578\,Mon}

\author[K. Pavlovski et al.]{K. Pavlovski$^{1}$\thanks{E-mail: pavlovski@phy.hr},
J.\ Southworth$^{2}$,
and E.\ Tamajo$^{3}$  \\
$^{1}$ Department of Physics, Faculty of Science, University of Zagreb, 10000 Zagreb, Croatia \\
$^{2}$ Astrophysics Group, Keele University, Staffordshire, ST5 5BG, UK \\
$^{3}$ University of Applied Science, 10410 Velika Gorica, Croatia \\
}
\date{Accepted XXX. Received YYY; in original form ZZZ}

\pubyear{2018}

\begin{document}
\label{firstpage}
\pagerange{\pageref{firstpage}--\pageref{lastpage}}
\maketitle

%%%%%%%%%%%%%%%%%%%%%%%%%%%%%%%%%%%%%%%%%%%%%%%%%%%%%%%%%%%%%%%%%%%%%%

% Abstract of the paper
\begin{abstract}
A prerequisite for probing theortical evolutionary models for high-mass stars 
is the determination of stellar physical properties with a high accuracy. 
We do this for three binary systems containing components with masses above 
10\Msun: V478\,Cyg, AH\,Cep and V453\,Cyg. New high-resolution and high-S/N 
\'echelle spectra were secured and analysed using spectral disentangling, 
yielding improved orbital elements and the individual spectra for the 
component stars. In conjuction with a re-analysis of archival light curves, 
the stellar masses and radii were measured to accuracies of 1.5--2.5\% and 
1.0--1.9\%, respectively. Detailed spectroscopic analysis then yielded 
atmospheric parameters and abundances for C, N, O, Mg and Si. Abundances were 
also determined for V578\,Mon. These results allowed a detailed comparison 
to the predictions of stellar evolutionary models. No star in our sample fits 
its position on the evolutionary track for its dynamical mass, leaving mass 
discrepancy (or alternatively overluminosity) an open problem in theoretical 
modelling. Moreover, the CNO abundances cluster around the initial values 
with mean values $\log (\textrm{N/C}) = -0.56 \pm 0.06$ and $\log (\textrm{N/O})
 = -1.01 \pm 0.06$. No trend in the CNO abundances for single OB stars are 
found for the eight high-mass stars analysed here, as well in an extended 
sample including literature results. This opens an important question on 
the role of binarity in terms of tides on damping internal mixing in stars 
residing in binary systems.
\end{abstract}

% Select between one and six entries from the list of approved keywords.
% Don't make up new ones.

%%%%%%%%%%%%%%%%%%%%%%%%%%%%%%%%%%%%%%%%%%%%%%%%%%%%%%%%%%%%%%%%%%%%%%

\section{Introduction}

Detached eclipsing binary stars (dEBs) are the primary source of fundamental 
stellar properties (mass $M$, radius $R$ and effective temperature \Teff). 
Direct determinations of mass, the most fundamental stellar quantity, can be 
obtained only from measurements of radial velocities (RVs) of the components 
in binary and multiple system, complemented with a determination of the 
inclination of the orbital plane either from analysis of light curves, 
or astrometric/interferometric measurements \citep[c.f.][]{hilditch_mono}. 
Dynamically determined stellar mass is a model-independent quantity. 
Assessing the fidelity of theoretical stellar evolutionary models is 
possible only with accurate empirical data \citep{pols, lastennet, willie_review}.

Despite the astrophysical importance of high mass stars in the critical 
evaluation of stellar theoretical predictions, only ten dEBs were known with 
physical properties measured to an accuracy and precision of 3\%, at the time 
of the most recent review paper \citep{willie_review}. Since then, this list 
has been expanded by only three systems 
(see DEBCat\footnote{\texttt{http://www.astro.keele.ac.uk/jkt/debcat/}}, 
\citealt{john_debs}), so the situation remains unsatisfactory. An important 
piece of analysis is also missing for many high-mass dEBs: a detailed 
spectroscopic analysis including atmospheric diagnostics and determination 
of the photospheric chemical composition for the individual components in a system.

Some steps have been undertaken in this direction \citep{pav_2005, pav_v453, 
pav_v380, mayer_hd, andrew_v380, andrew_sigma, andrew_spica, martins_mahy}, 
but the small sample size prevents definitive conclusions. However, the results 
of these studies have clearly confirmed the disagreement between empirical data 
and theoretical evolutionary models: high-mass stars are in general more luminous 
than predicted by theoretical calculations for their measured dynamical masses. 
This effect was discovered and named the `mass discrepancy' by \citet{herrero}, 
who compared spectroscopically determined masses to evolutionary tracks for single
 high-mass stars. The inclusion of rotation into theoretical models was 
a promising solution, as stellar rotation could induce some extra missing of 
stellar material, and moreover lift nuclear-processed material up to the surface 
layers \citep{maeder_araa, langer_araa}. However, observational studies have not 
completely supported this concept \citep{hunter_letter, hunter_paper}, opening 
more problems than were solved. Whilst changes in the carbon, nitrogen and oxygen 
(C, N and O) abundances in the photospheres of high-mass single stars were proved 
with dedicated accurate observations \citep{przybilla_cno, fernanda_standard, maeder_cno}, 
the role of rotation remains unclear and possibly unhelpful \citep{conny_nitro}.
 The observational studies of  highly rotating  stars with the masses larger
than 20 \Msun \citep{cazorla2017a,cazorla2017b, markova} are also inconclusive
for the role of rotational mixing in changing the abundance pattern for the
CNO elements in stellar atmospheres. Promising direction in understanding internal
structure and mixing profiles for  high mass stars is coming through asteroseismic probing
as was shown in the calculations by \citet{pedersen}. The authors showed that a combination
of gravity-mode oscillations and surface nitrogen abundance is extremely sensitive probe 
for constraining not only the amount but also the profile of interior mixing in
intermediate- and high-mass stars.

Recently, \citet{garland_tarantula} examined a sample of B-type multiple systems 
in the VLT/FLAMES Tarantula Survey. Due to observational constraints the authors 
limited their sample to the subset of 33 systems with relatively small projected 
rotational velocities ($v\sin i \leq 80$\kms). Their abundance analysis revealed 
primary stars with a whole range of N enhancement up to a normal abundance, thus 
corroborating the N abundances found for single B-type stars in the LMC. The 
majority of the stars studied in their sample were identified as binary systems 
from RV variability and/or detection of a secondary component in the spectrum. 
Their results suffer from unknown contribution and flux dilution due to the 
presence of the secondary star, which would affect the measurement of the 
atmospheric parameters and abundances. Since abundance changes are sensitive 
to bulk metallicity, binaries in the Local Group of galaxies like the LMC and 
SMC are important for further investigation.

\begin{table*}
\centering
\caption{\label{tab:sample}
Basic characteristics of binary systems studied in the present work. References to
quoted values are given in Section~\ref{sec:sample} which describes the binaries 
in our sample.}
\begin{tabular}{lccccccc} \hline
Binary   & HD      & Orbital    & $V_{\rm max}$ & Spectral            & Age   & Cluster    & Apsidal               \\
system   & number  & period [d] & [mag]         & types               & [Myr] & membership & period [yr]           \\
\hline
V478~Cyg & 121721  & 2.88       & 8.63          & O9.5\,V + 09.5\,V   & 10    & Cyg\,OB1   & $27.1 \pm 0.5$        \\
AH~Cep   & \;99272 & 1.77       & 6.88          & B0.5\,Vn + B0.5\,Vn & 3.2   &            &                       \\
V453~Cyg & 126721  & 3.89       & 8.28          & B0.4\,IV + B0.7\,IV & 6.9   & Cyg\,OB2   & $66.4 \pm 1.8$        \\
V578~Mon & \;53121 & 2.41       & 8.55          & B1\,V + B2\,V       & 2.2   & NGC\,2244  & $33.48^{+0.10}_{-0.06}$ \\
\hline
\end{tabular}
\end{table*}

Empirical constraints on these processes remain hard to come by, despite a steady 
improvement in observational techniques and capabilities 
\citep[see][]{hilditch_2004}. In the current series of papers we aim to calibrate 
the abundance patterns and chemical evolution of a sample of high-mass stars 
by analysing dEBs. These are vital for obtaining observational constraints on 
the structure and evolution of high-mass stars, and are the primary source of 
directly measured stellar properties \citep{Andersen91aarv}. Chemical abundances 
are difficult to determine from the spectra of high-mass dEBs for several reasons.
 Firstly, they tend to display only a small number of spectral lines. Secondly, 
the often high $v\sin i$ values of these objects means the lines are wide and 
shallow. Thirdly, in dEBs the spectral lines from the two components interfere 
with each other (`line blending').

In a seminal work, \citet{klaus_spd} introduced the technique of {\em spectral
 disentangling} (\spd), by which {\em individual} spectra of the component 
stars of a double-lined spectroscopic binary can be deduced from observations 
covering a range of orbital phases. \spd\ can be used to measure spectroscopic 
orbits which are not affected by line blending \citep[see][]{jkt_dwcar}. 
The resulting disentangled spectra also have a much higher S/N than the original 
observations, making them very useful for chemical abundance analysis. 
As a bonus, the strong degeneracy between \Teff\ and surface gravity (\logg) is 
not a problem for dEBs because \logg\ can be measured directly and to high 
accuracy (0.01\,dex or better). A detailed investigation of these possibilities 
is given by \citet[][hereafter PH05]{pav_2005}.

%%%%%%%%%%%%%%%%%%%%%%%%%%%%%%%%%%%%%%%%%%%%%%%%%%%%%%%%%%%%%%%%%%%%%%

\section{Sample} \label{sec:sample}

Four main-sequence (MS) binary systems have been selected for study, based partly 
on the availability of high-resolution spectra. These are V478\,Cyg, AH\,Cep, 
V453\,Cyg and V578\,Mon, all composed of components of late-O or early-B spectral 
type. Basic information on these systems is given in Table~\ref{tab:sample}. 
The component stars have \Teff s between 25\,000 and 32\,000 K, and \logg s 
from 3.7 to 4.2. The primary star of V453\,Cyg is near the terminal-age MS, 
the components of V478\,Cyg are roughly halfway through their MS lifetimes, 
and the components of AH\,Cep are early in their MS lifetimes. We now introduce
 each of these systems.

\subsection{V478 Cygni}

V478\,Cyg is the least-studied system considered here. This is surprising 
becasuse it contains two late-O stars with masses slightly above 16\Msun, 
so is very useful for constraining the evolution of high-mass stars. The only 
modern spectroscopic study is due to \citet{popper_1991}, who secured observations
 at the Lick Observatory. The cross-correlation technique was applied to derive 
RVs using He\,{\sc i} lines, based on digitised versions of the photographic-plate
 spectra. The RVs were complemented with extensive photometric observations 
\citep{popper_1991, sezer} and the properties of the components were derived. 
This revealed it to be a binary system of two almost identical components.

\citet{mossakovskaya} found apsidal motion from photometric observations, 
and estimated an age of about 4\,Myr from a comparison of the observed masses 
and radii to theoretical models. \citet{wolf_v478} presented additional times 
of minimum light and determined an apsidal period of $U = 27.1 \pm 0.5$\,yr 
and an eccentricity of $e = 0.0158 \pm 0.0007$ \citet{claret_apsidal} analysed 
the apsidal motion for a sample of well-determined binaries, including V478\,Cyg, 
and found an excellent agreement between the observations and stellar models.

\citet{zakirov_v478} presented and analysed extensive $UBVR$ photometry of 
V478\,Cyg. \citet{zakirov_dol42} studied the membership probability of this 
binary system in the open cluster Dolidze 42 using multicolour photometry. 
They found that it is a background object, but is probably a member of the 
OB association Cyg\,OB1 with the distance of 1.04\,kpc and an age of about 
10\,Myr.

\subsection{AH Cephei}

AH\,Cep is a short-period dEB composed of two high-mass early-B stars 
\citep{bell_ahcep,holmgren_ahcep}. Modern photometric solutions 
\citep{bell_ahcep, horst_ahcep, harvig_ahcep} have shown that the system 
is detached and at the beginning of its MS lifetime with an age of about 6\,Myr. 
Unfortunately, it is not a member of any known stellar cluster or association, 
so an independent estimate of the age is not available.

Some ambiguity is left in the photometric solution primarily because of the 
uncertainties in the component \Teff s. The rather low orbital inclination, 
which causes the eclipses to be partial and shallow, has contributed additional 
uncertainties especially when the light ratio of the two stars is not fixed via 
spectroscopic analysis (see Section~\ref{sec:orbits:spd}).

There are indications for the presence of a third body in the system, on the 
basis of the period study by \citet{mayer_ibvs} and third light in light curve 
solutions \citep{horst_ahcep}. However, the analysis of times of minima by 
\citet{bell_ahcep} did not support the claimed light-time effect. This issue 
has been discussed by \citet{harvig_ahcep} and \citet{kim_ahcep}. These authors 
also suggested that a fourth body could exist. Also uncorroborated is the claim 
by \citet{horst_ahcep} that long-term changes in the amplitude of the light 
curve exist, which these authors explained by the effect of changing inclination.

The $v\sin i$ values of the components are relatively high (185\kms, 
\citealt{holmgren_ahcep}), so the lines available for RV measurement are rather 
broad and/or shallow, and difficult to measure. This might explain the poor 
agreement between different datasets. Modern spectroscopic studies have yielded 
masses of the components of $M_{\rm A} = 18.0 \pm 1.2$\Msun\ plus  $M_{\rm B} = 
15.9 \pm 1.1$\Msun\ \citep{bell_ahcep}, $M_{\rm A} = 15.4 \pm 0.2$\Msun\ plus 
 $M_{\rm B} = 13.6 \pm 0.2$\Msun\ \citep{holmgren_ahcep}, and $M_{\rm A} = 16.1 
\pm 0.6$\Msun\ plus $M_{\rm B} = 13.3 \pm 0.4$\Msun\ 
\citep{massey}\footnote{There is a numerical error in the calculation of the 
uncertainties in \citet{massey} as they used an inclination of $i = 69.2^{\rm o} 
 \pm 12^{\rm o}$ instead of the value of $i = 69.21 \pm 0.12$ given by 
\citet{bell_ahcep}.}.

\citet{Ignace} analysed {\em Chandra} observations and found X-ray variability. 
They also studied CW\,Cep, a binary system with very similar physical properties 
to AH\,Cep. CW\,Cep was not detected as X-ray source, so they excluded embedded 
wind shocks or binary wind collision as the source of the X-rays in AH\,Cep. 
The X-rays were instead attributed to magnetism in one or both components of 
AH\,Cep.

\subsection{V453 Cygni}

V453\,Cyg is the only totally-eclipsing binary in our sample. Its primary 
(more massive, and hotter) component has passed more than half its MS lifetime, 
and is evolutionarily the most advanced among the stars in our sample. 
In a comprehensive study of V453\,Cyg \citet{jkt_v453} analysed new grating 
spectra and the light curves from \citet{cohen}, obtaining masses of $M_{\rm A} 
= 14.36 \pm 0.20$\Msun\ and $M_{\rm B} = 11.11 \pm 0.13$\Msun, and radii of 
$R_{\rm A} = 8.55 \pm 0.06$\Rsun\ and $M_{\rm B} = 5.49 \pm 0.06$\Rsun. The 
high precision of the radius determination was greatly helped by the deep and 
total eclipses. An apsidal period of $U = 66.4 \pm 1.8$\,yr was found.

The same set of grating spectra, centred on the H$\gamma$ line, were complemented 
with grating spectra obtained by \citet{klaus_spd} and further analysed by 
\citet{pav_v453}. They applied \spd\ to isolate the individual spectra of 
the components. The spectroscopic analysis yielded $T_{\rm eff, A} = 27\,900 
\pm 400$\,K and $T_{\rm eff, B} = 26\,200 \pm 500$\,K. Abundances for both 
components were determined for the first time. The abundances were found to
 be typical for B-type stars, whilst no indication was found for the presence 
of CNO-processed material in the photospheres of the component stars.

\subsection{V578 Monocerotis}

V578\,Mon is composed of two early-B stars. It is a member of the young open 
cluster NGC\,2244, which is itself embedded in the Rosette Nebula (NGC\,2237, 
NGC\,2246). This star-forming region is situated inside the older Mon OB2 
association (see \citealt{Lada} and references therein). \citet{Heiser_1977} 
found V578\,Mon to show fast apsidal motion. The first comprehensive analysis 
of photometric and spectroscopic observations for this system was accomplished 
by \citet{hensberge_2000}, based on $uvby$ photometric measurements and 
high-resolution \'echelle spectra. Application of \spd\ made possible the 
determination of the orbital and stellar quantities with a high accuracy.

\citet{Garcia_2011} studied apsidal motion for this system, and improved 
the measurement of its orbital eccentricity. \citet{Garcia_2014} reanalysed 
all available light curves, complemented with new high-resolution spectra 
secured with the {\sc hermes} spectrograph at the Mercator telescope on La 
Palma, Spain. Their results were fully compatible with those of 
\citet{hensberge_2000}, with a further improvement in the accuracy. Hence 
the uncertainties in mass are around 0.5\%, in radius around 1\%, and in 
\Teff\ around 1.7\%. They also found $U = 33.48^{+0.10}_{-0.06}$\,yr. These 
results make V578\,Mon a benchmark dEB among the high-mass binaries, and 
allows a strict test of stellar evolutionary models \citep{Garcia_2014}.

V578\,Mon was detected in the high-spatial-resolution X-ray study of NGC\,2244 
using {\em Chandra} \citep{Wang}. The authors attributed the X-ray spectra of 
the Rosette Nebula OB stars to the standard model of small-scale shocks in the 
inner wind of these high-mass stars. Since V578\,Mon is an eclipsing system it 
offers a rare opportunity where X-ray eclipses can be used to constrain the 
emitting geometry. However, to our knowledge no further X-ray study of this 
binary system has been undertaken.

%%%%%%%%%%%%%%%%%%%%%%%%%%%%%%%%%%%%%%%%%%%%%%%%%%%%%%%%%%%%%%%%%%%%%%

\begin{figure*}
\begin{tabular}{cc}
\includegraphics[width=81mm]{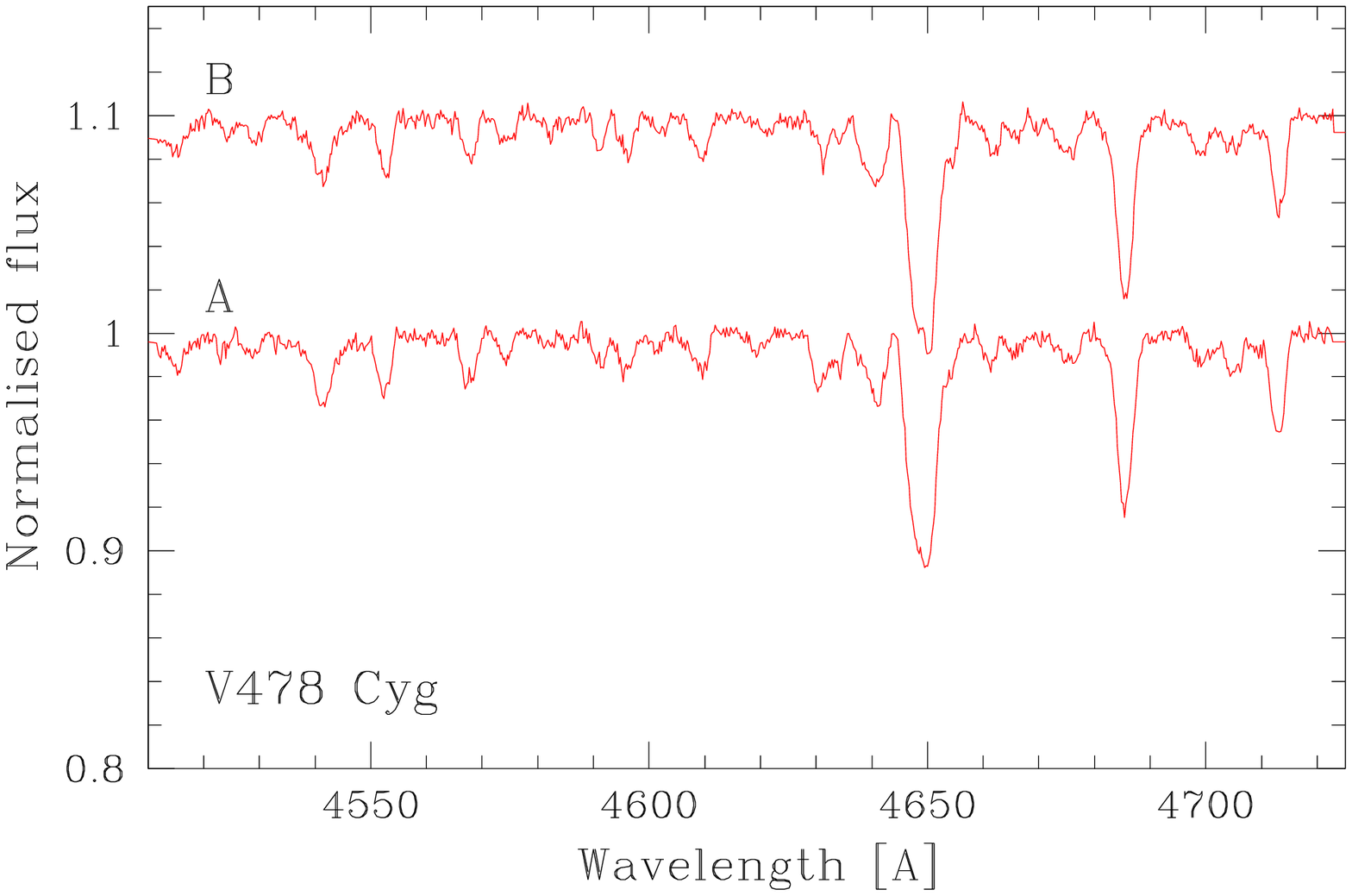} &
\includegraphics[width=81mm]{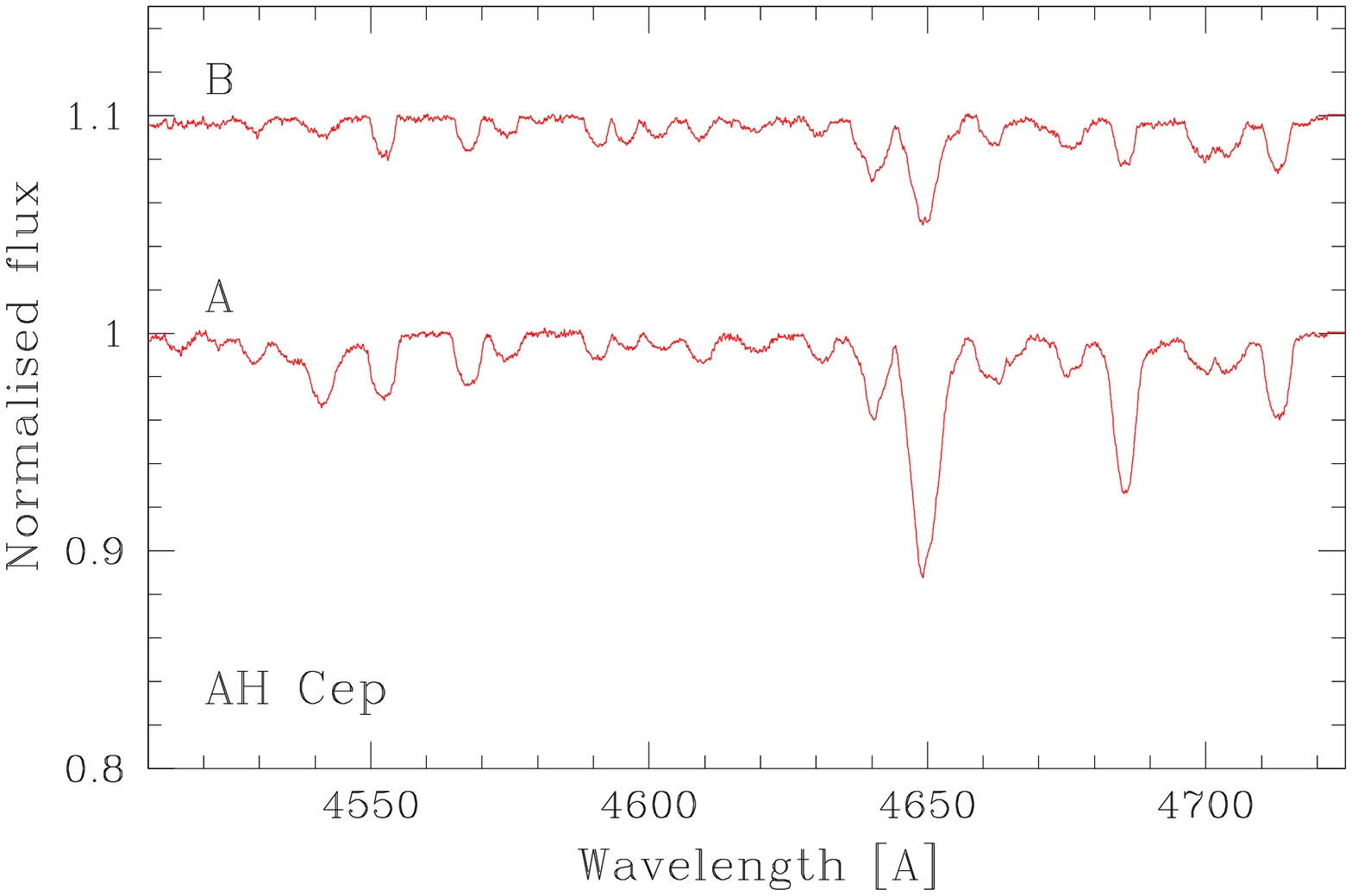} \\
\includegraphics[width=81mm]{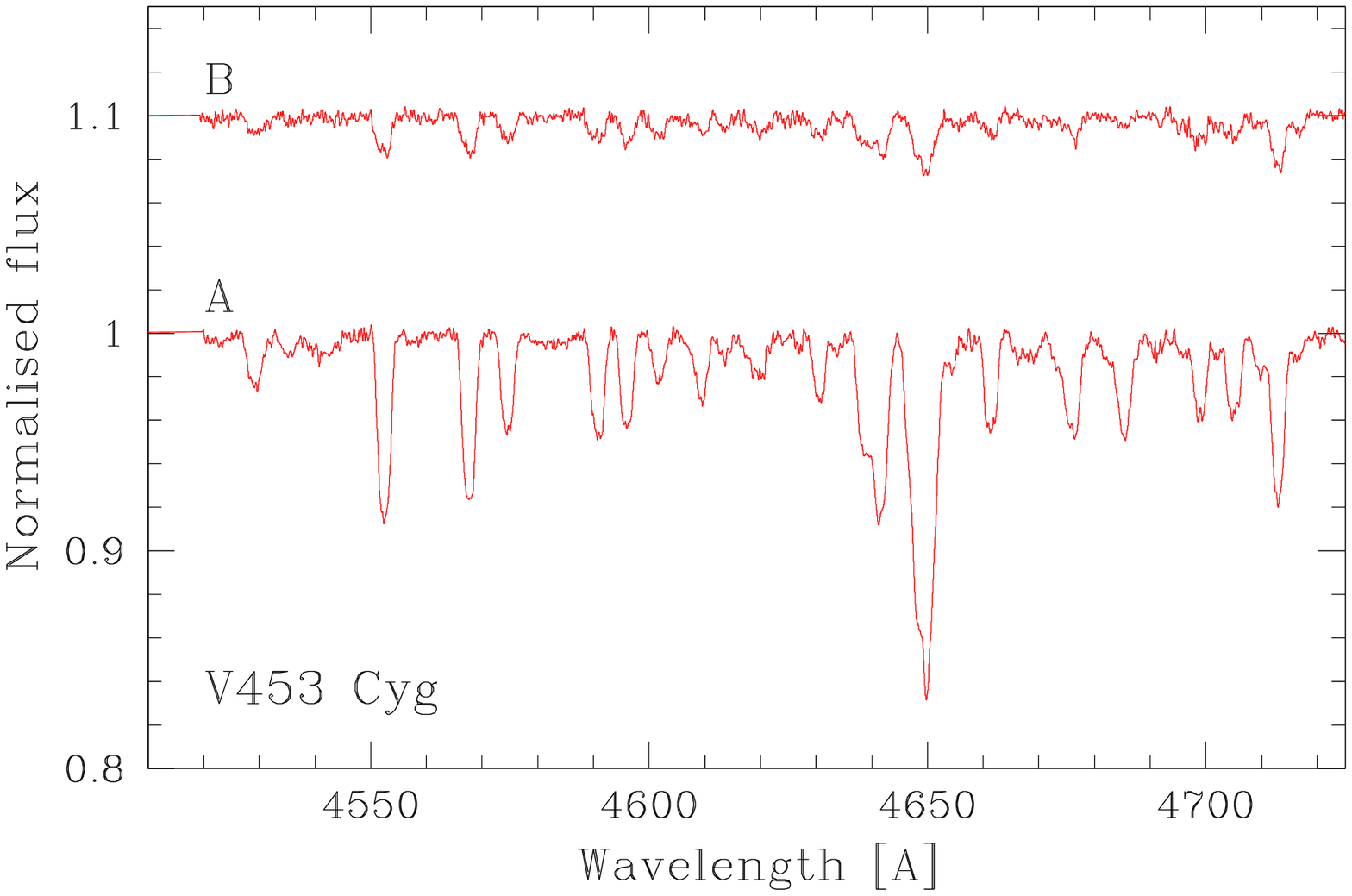} &
\includegraphics[width=81mm]{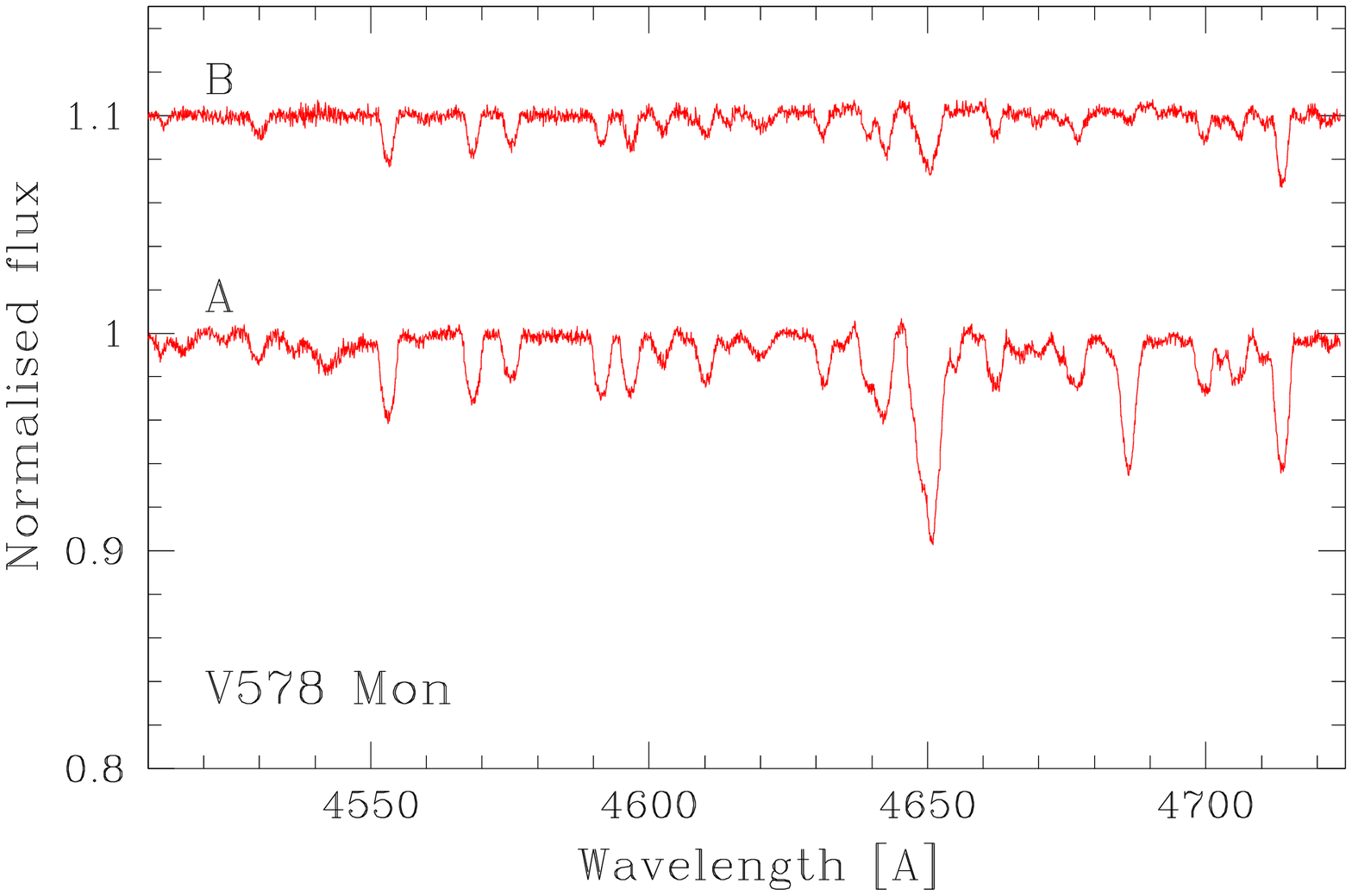} \\
\end{tabular}
\caption{\label{fig:plotspe}
Portions of the disentangled spectra of the stars (labelled) studied in this work.}
\end{figure*}

\section{Spectroscopic data}

New spectroscopic observations have been secured with three high-resolution 
\'echelle spectrographs.

A total of 13 spectra for V478\,Cyg were aquired in June 2013 with the Hamilton 
Echelle Spectograph \citep{hamilton} on the Shane 3\,m telescope at Lick 
Observatory, USA. The spectra cover 3550--7280\,{\AA} over 89 \'echelle orders 
and with resolving power $R = 35\,000$. The signal to noise ratios (S/N) in 
the visual range from 53 to 92 and average 70.

AH\,Cep was observed using the Fibre-fed Echelle Spectrograph (FIES, 
\citealt{fies}) aT the 2.56\,m Nordic Optical Telescope at La Palma, Spain. 
26 spectra were obtained in November
2006 and October 2007. The spectra cover 3440--9700\,\AA\ over 79 \'echelle 
orders with $R = 40\,000$. The S/N values range from 230 to 355 and average 300.

21 spectra of V453\,Cyg were obtained in August 2008 using the Fiber Optics 
Cassegrain Echelle Spectrograph (FOCES, \citealt{foces}) on the 2.2\,m telescope 
at Calar Alto, Spain. They cover 3850--7340\,\AA\ at $R = 45\,000$, and have 
S/N values from 80 to 110 with an average of 97.

The spectra for V578\,Mon \citep{Garcia_2014} were acquired with the High 
Efficiency and high Resolution Mercator Echelle Spectrograph (HERMES, 
\citealt{hermes}) on the 1.2\,m Mercator telescope at La Palma, Spain. The 
33 spectra cover 3700--9100\,\AA\ at $R = 85\,000$ and with S/N values from 70 
to 110.

In each case, wavelength calibration was performed using thorium-argon exposures, 
and flat-fields were obtained using a halogen lamp. The FIES and FOCES spectra 
were bias-subtracted, flat-fielded and extracted with the 
{\sc iraf}\footnote{{\sc iraf} is distributed by the National Optical Astronomy 
Observatory, which are operated by the Association of the Universities for 
Research in Astronomy, Inc., under cooperative agreement with the NSF.} 
\'echelle package routines. The Hamilton spectra were reduced with the 
standard reduction package for this instrument. Normalisation and merging 
of the orders was performed with great care, using programs developed by 
ourselves \citep{kolbas_algol}, to ensure that these steps did not cause 
any systematic errors in the resulting spectra.

\begin{table*}
\centering
\caption{\label{tab:orbit}
Parameters of the spectroscopic orbits for V478\,Cyg, AH\,Cep and V453\,Cyg 
determined by the \spd\ method in this work. Throughout the analysis the period 
was kept fixed to the value given the Table. }
\begin{tabular}{lccccccc} \hline
Binary   & Period & $T_{\rm peri}$ & $e$  & $\omega$  &  $K_{\rm A}$ & $K_{\rm B}$ & $q$   \\
system   & [d]   & [d]             &      & [deg]     &  [km\,s$^1$] & [km\,s$^1$] &    \\
\hline
V478\,Cyg   & 2.8808994 & 2456222.23 $\pm$ 0.19 & 0.021 $\pm$ 0.005 & 148.1 $\pm$ 21.3 & 225.9 $\pm$ 2.3 & 231.5 $\pm$ 2.6 & 0.976 $\pm$ 0.011 \\
AH\,Cep     & 1.7747420 &    -  &        0.0          &  -               & 234.9 $\pm$ 1.1 & 276.9 $\pm$ 1.4 & 0.848 $\pm$ 0.009 \\
V453\,Cyg   & 3.889825  & 2445272.31 $\pm$ 0.16 & 0.022 $\pm$ 0.003 & 25.1 $\pm$ 11.7   & 175.2 $\pm$ 1.3 & 220.2 $\pm$ 1.6 & 0.795 $\pm$ 0.007 \\
\hline \\
\end{tabular}
\end{table*}

%%%%%%%%%%%%%%%%%%%%%%%%%%%%%%%%%%%%%%%%%%%%%%%%%%%%%%%%%%%%%%%%%%%%%%

\section{Spectroscopic orbits}  \label{sec:orbits}

\subsection{Spectral disentangling} \label{sec:orbits:spd}

Our analysis follows the methods introduced by \citet{hensberge_2000} and 
PH05 in their studies of V578\,Mon. This approach involves reconstruction 
of individual component spectra from the observed composite spectra using 
\spd, which then enables a detailed abundance analysis using the same methods 
as for single stars \citep{hensberge_prague, pav_brno, pav_iau}. \spd\ also 
gives the velocity amplitudes of the two stars which, together with analysis 
of light curves, yields the physical properties of the stars. The abundance 
analysis then can benefit from the availability of the  precise and accurate 
surface gravity measurements from the known masses and radii of the stars.

The spectra of OB binaries are dominated by pressure-broadened H and He lines, 
and usually rotationally broadened metal lines. Although metal lines are not 
numerous, and are due to only a few species at these high \Teff s, line blending 
is severe and further enhanced by Doppler shifts in the course of the orbital 
cycle. This is the principal reason the masses and atmospheric properties for 
most high-mass stars are not derived with an accuracy usually achieved for 
low-mass stars. \spd\ enables reconstruction of individual spectra of the 
components from a time-series of composite spectra, while simultaneously 
solving for the optimal orbital elements. There is no need for template 
spectra which also means biases due to template mismatch are avoided 
\citep{hensberge_prague}.

We performed \spd\ in the Fourier domain \citep{hadrava_1995} using the code 
{\sc fdbinary}\footnote{\tt http://sail.zpf.fer.hr/fdbinary/} \citep{sasa_spd}. 
One advantage of Fourier disentangling is that only modest computing resources 
are needed, so long or multiple spectral segments can be disentangled. Some 
practical issues for \spd\ are discussed in \citet{pav_brno}. A regular 
distribution of observations through the orbital cycle is important, as was 
shown numerically by \citet{hensberge_2008}. Otherwise, a spurious pattern 
in disentangled spectra can appear, diminishing the quality of reconstructed 
spectra and affecting the reliability of the orbital solution.

In {\sc fdbinary} optimisation is performed with the simplex routine 
\citep{press1992}. We usually performed 100 runs, each with 1000 iterations, 
examining a relatively wide parameter space around the initial set of orbital 
parameters. In most cases, with high-S/N spectra well distributed in the orbital
 phases, convergence was achieved quickly.

\subsection{Error calculations for the orbital parameters in spectral 
disentangling with bootstrapping}

\spd\ allows the disentangled spectra of the component stars to be determined 
simultanously with the orbital elements of the binary system. Propagation of 
uncertainties through this process is difficult so must be tackled numerically. 
The {\sc fdbinary} code includes an implementation of the jackknife method 
\citep{Tukey}, which is an early example of resampling methods with limited 
statistical validity. We have therefore quantified the uncertainties of the 
orbital elements from \spd\ using bootstrapping \citep{Efron}, a versatile 
generalisation of jackknife. Bootstrapping is an efficient tool for 
constraining confidence intervals and calculating standard errors 
\citep{Lupton,Ivezic}.

Our approach is to generate a large number of spectral datasets by sampling 
from the original spectra with replacement. Whilst for $N$ observed spectra 
the jackknife method allows only $N-1$ unique samples to be obtained, 
bootstrapping allows for up to $N!$ samples although in practise of order 
1000 samples are sufficient. Each of the spectral datasets are run through 
{\sc fdbinary} in order to determine a distribution of values for each orbital 
parameter of interest. From these distributions we calculate the standard 
deviation of the values for each parameter and adopt this as its uncertainty. 
{\sc fdbinary} is well suited to this approach because its use of the Fourier 
domain makes it relatively fast, although it remains computationally demanding 
for eccentric orbits.

\begin{figure*}
\begin{tabular}{cc}
\includegraphics[width=80mm]{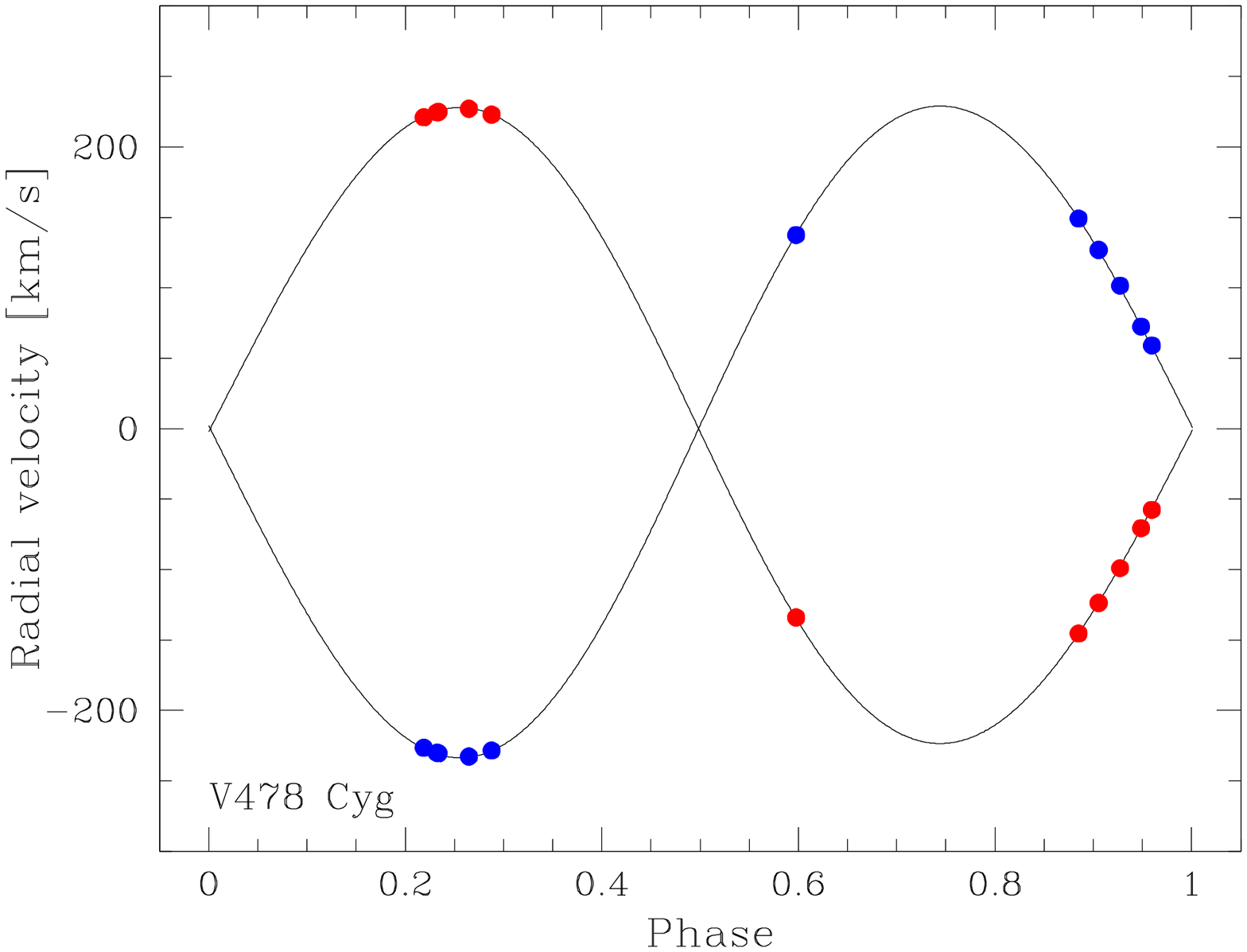} &
\includegraphics[width=80mm]{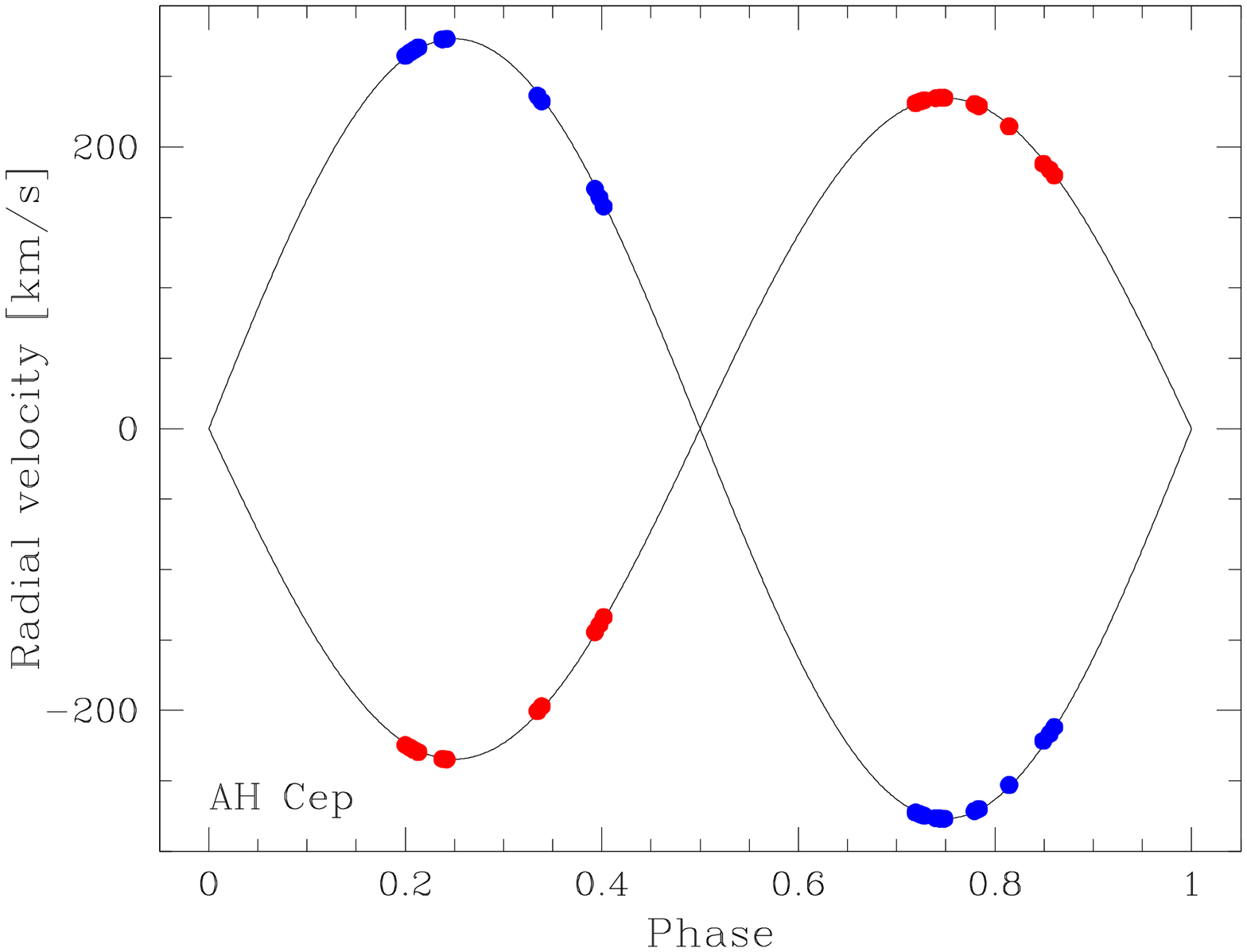} \\
\includegraphics[width=80mm]{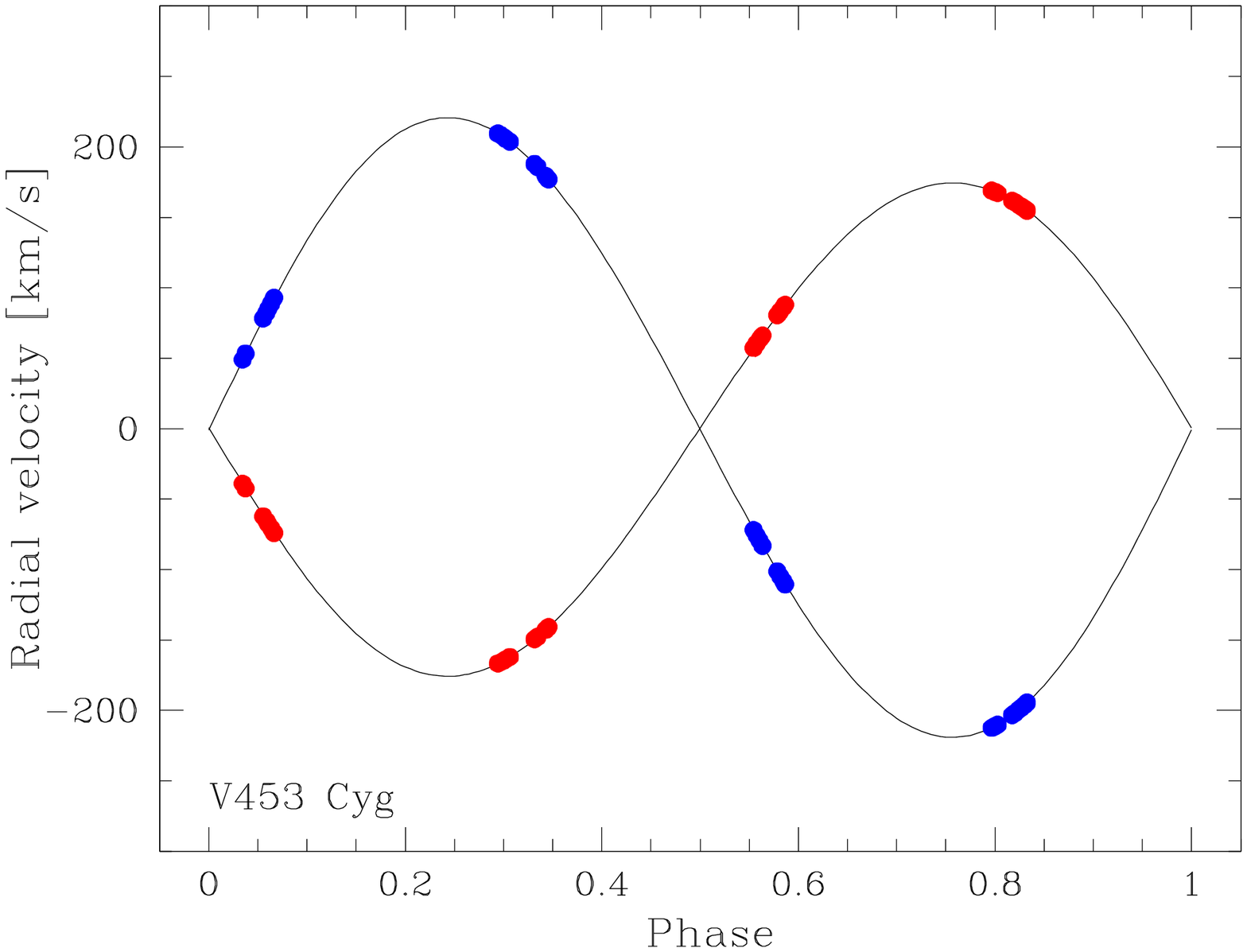} &
\includegraphics[width=80mm]{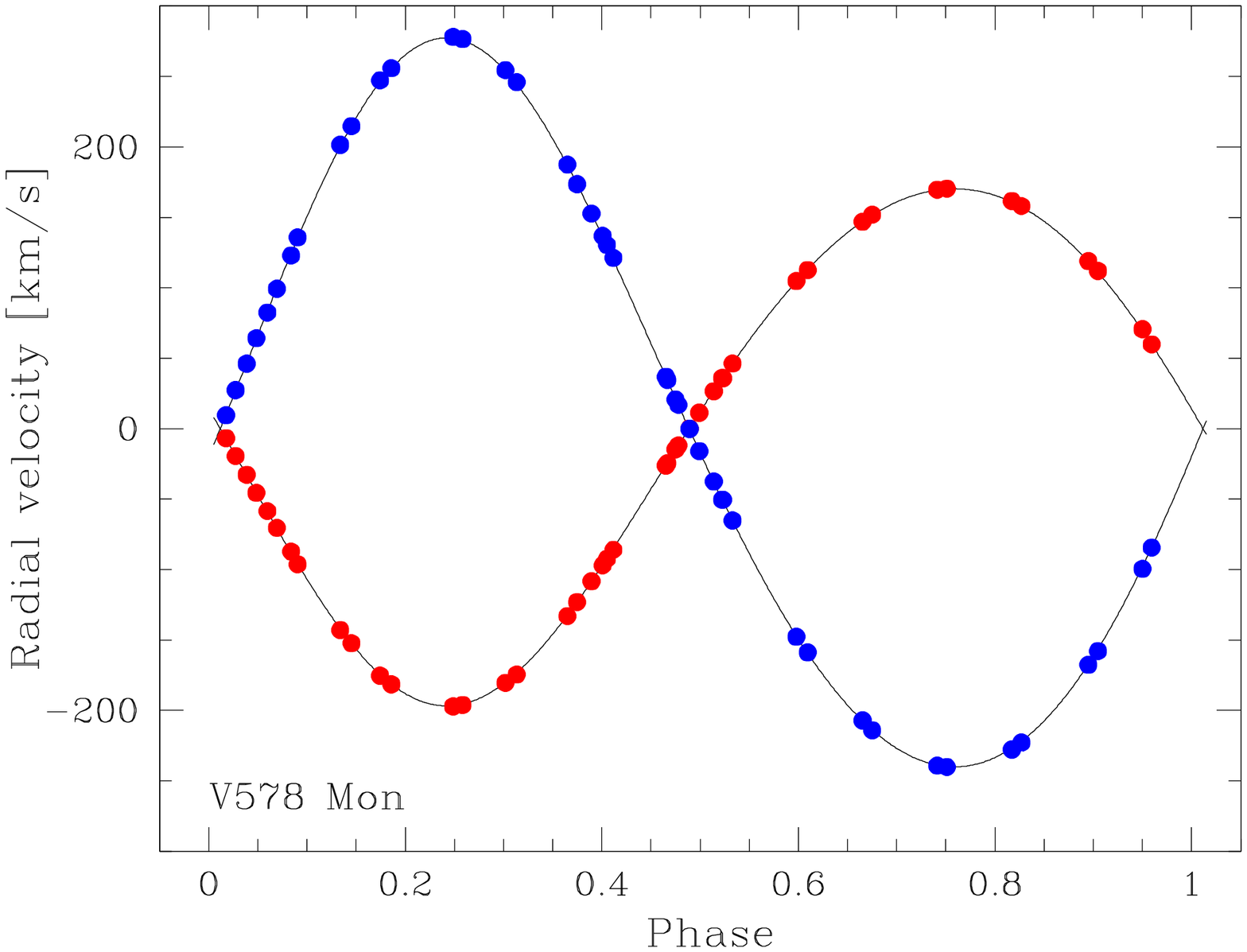} \\
\end{tabular}
\caption{\label{fig:plotorbit}
Visualisation of the spectroscopic orbits of our targets. The best-fitting orbits
 are shown with black lines and the RVs of the stars at the times of observation 
with blue and red circles. Note that these are not measured
RVs: these plots are here to indicate the orbital phase coverage and range of 
RVs sampled for each system.}
\end{figure*}

\subsection{Orbit determination}

In \spd\ the orbital parameters are determined simultaneously with a 
reconstruction of the components' individual spectra using a time-series of 
the observed spectra after careful continuum normalisation \citep{kolbas_algol}. 
First, we searched for the orbital solution by performing \spd\ on relatively 
short spectral segments of about 100--300\,\AA\ width, centred on the prominent 
\ion{H}{i} and \ion{He}{i} lines. Since the spectra of all binary systems 
analysed here are affected by diffuse interstellar bands (DIBs) and interstellar 
lines, special care was taken to mask them. The red parts of the spectra were 
also not used because of variable telluric lines which cannot be handled with 
the present version of {\sc fdbinary}.

In the final run for the orbital solution we performed \spd\ in a large spectral 
range covering about 4000 to 6000 {\AA}, masking the broad Balmer lines and DIBs. 
Thus, the orbital parameters are derived from \ion{He}{i}, \ion{He}{ii}, and 
numerous metal lines, of which the lines of \ion{He}{i}, and \ion{He}{ii} are 
the strongest in almost all spectra of the individual components. The Balmer 
lines were excluded from this analysis because Doppler shifts are much smaller 
than intrinsic width of the lines, and some unavoidable systematic uncertainties 
remain in placing the continuum level for these broad lines.

\subsubsection{V478 Cyg}

Our 11 spectra are not ideally distributed in orbital phase.  The first 
quadrature is covered well, but the second is not (see Fig.\,\ref{fig:plotorbit}).
 We fitted for the parameters of an eccentric orbit, since indications of orbital
 eccentricity were found previously \citep{popper_1981}. The results are given 
in Table~\ref{tab:orbit}. The errors were calculated via bootstrapping. We 
confirm the finding of \citet{popper_1991} that the masses of the components 
are very similar. The mass ratio determined with \spd, $q = 0.976 \pm 0.013$ 
is in excellent agreement with that found by \citet{popper_1991}, $q = 0.979 
\pm 0.012$  and within 1-$\sigma$ uncertainty corroborates with the 
result of \citet{martins_mahy} who derived $q = 0.956 \pm 0.033$. Latter
result is also based on high-resolution CCD spectra, and using {\sc spd}
method for determination of the spectroscopic orbit.

We find that the velocity amplitudes $K_{\rm A}$, and $K_{\rm B}$ determined 
via \spd\ are systematically lower than those measured using cross-correlation 
\citep{popper_1991}. The difference is about 6\kms\ and means our measured 
masses are lower than those found by \citet{popper_1991}. \citeauthor{popper_1991}
 introduced an empirical correction factor calibrated in cross-correlation of 
standard RV stars. If uncorrected values are considered, then the differences 
in $K_{\rm A}$ and $K_{\rm B}$ between their and our result amount to about 13\kms.
 This is true also for the measurements performed by \citet{martins_mahy}
who obtained $K_{\rm A} = 222.1 \pm 5.4$ \kms, and $K_{\rm B} = 232.3 \pm 5.7$
\kms, but setting the eccentricity $e = 0$, as \citet{popper_1991} did, too.

V478\,Cyg comprises two almost identical components, as is shown by the almost 
indistinguishable disentangled spectra. In subsequent spectroscopic analysis 
we found a \Teff\ difference between the stars of $\Delta T \sim 300$\,K. The 
$v\sin i$ values are high for both stars, at $\sim$125\kms, making the lines 
broad and blended. We therefore ascribe the discrepancy with the analysis of 
\citet{popper_1991} to line blending affecting the cross-correlation solution
 of the photographic spectra with medium spectral resolution.

\subsubsection{AH Cep}

The orbital phase distribution of our 23 spectra is shown in 
Fig.~\ref{fig:plotorbit}. Whilst the observations do not cover a whole orbital 
cycle, both quadratures are well covered. This, and the fact that the orbit is 
circular, helps to measure $K_{\rm A}$ and $K_{\rm b}$ with reasonable 
uncertainties. We find $K_{\rm A} = 234.9 \pm 1.1$\kms\ and $K_{\rm B} = 
276.9 \pm 1.4$\kms\ (see Table~\ref{tab:orbit}), giving a mass ratio of 
$q = 0.848 \pm 0.009$.

Our new results fall in between those of previous studies. The mass ratios 
obtained by \citet{bell_ahcep} and \citet{holmgren_ahcep} are the same, 
$q = 0.88 \pm 0.04$, and $q=0.88 \pm 0.01$, respectively. However, the 
velocity amplitudes from these two works are very different. \citet{bell_ahcep} 
combined photographic and digital spectra, and found $K_{\rm A} = 249 \pm 8$\kms\ 
and $K_{\rm B} = 283 \pm 8$\kms. In their pioneering study using cross-correlation
 of Reticon spectra, \citet{holmgren_ahcep} arrived at substantially smaller 
values of $K_{\rm A} = 237 \pm 2$\kms\ and $K_{\rm B} = 269 \pm 2$\kms. In a 
third spectroscopic study of AH\,Cep \citep{massey} CCD spectra were obtained 
with a better spectral resolution than previously. Double Gaussians were fitted 
to the best double lines in the spectra, giving $K_{\rm A} = 230.0 \pm 3.2$\kms\ 
and $K_{\rm B} = 277.6 \pm 4.4$\kms. The velocity amplitudes and mass ratio, 
$q = 0.829 \pm 0.018$, are within 2$\sigma$ of our values. The previous 
inconsistencies can therefore be attributed to the variety of methods and 
observational data employed in the previous studies of AH\,Cep, as well as 
the intrinsic difficulty of the system.  The most recent spectroscopic study
\citep{martins_mahy} is not helpful in resolving possible ambiguity in
the spectroscopic orbit due to large uncertainties in derived semiamplitudes,
$K_A = 236.5\pm7.3$ \kms, and $K_B = 267.6\pm8.3$ \kms, with $q = 0.884 \pm 0.043$.

Both stars are rotating quickly, with $v \sin i \sim 170$--$180$\kms, which 
makes the measurements of the RVs extremely difficult. In an extensive 
intercomparison of different measuring techniques for a similar system, 
\citet{jkt_dwcar} found that \spd\ gave the most internally consistent results, 
followed by double-Gaussian fitting. This supports the agreement between our 
orbital elements and those found by \citet{massey}; our analysis also benefited 
from using high-resolution and high-S/N \'echelle spectra.

\begin{figure}
%\begin{tabular}{ccc}
\includegraphics[width=80mm]{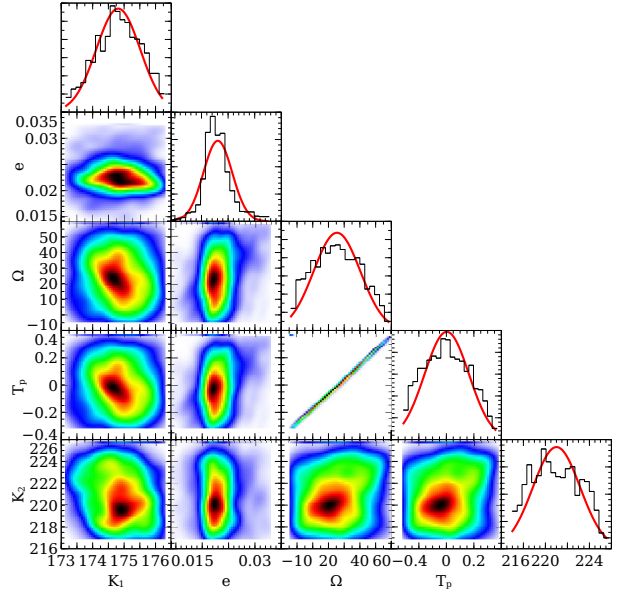}
%\end{tabular}
\caption{\label{fig:plotboot}
Example corner plot showing the distribution of values of the orbital parameters 
from 2000 boostrapping simulations for V453\,Cyg, determined using \spd. 
The 1$\sigma$ confidence levels correspond to the yellow shading.}
\end{figure}

\subsubsection{V453 Cyg}

Our 33 new spectra cover the range of RVs of the two stars fairly well 
(Fig.\,\ref{fig:plotorbit}). The orbital parameters were optimised as before, 
using all spectra and the spectral range 4000--6000\,\AA\ with masks on broad 
Balmer and interstellar lines. The uncertainties were calculated with 2000 
bootstrap samples. The orbit of V453\,Cyg is eccentric, and all parameters 
were initially set free in the calculations.

The orbital solution gives $K_{\rm A} = 175.2 \pm 0.7$\kms\ and $K_{\rm B} = 
220.2 \pm 1.6$\kms, which yields $q = 0.795 \pm 0.007$ (Table~\ref{tab:orbit}). 
In our previous studies of V453\,Cyg the orbital solution was determined from 
grating spectra \citep{jkt_v453,pav_v453} and we found $K_{\rm A}$ values of 
172.5--174.1\kms, and $K_{\rm B}$ values of 216.2--224.6\kms. Our new solution, 
based on \'echelle spectra, falls between these values. The eccentricity 
$e = 0.022 \pm 0.002$ is in good agreement with that derived in \citet{jkt_v453}
 from apsidal motion, and within 1$\sigma$ of that found by \citet{pav_v453}.

Orbital solutions for V453\,Cyg have also been published by \citet{popper_1991}, 
\citet{klaus_spd} and \citet{massey}. The methods used by \citet{popper_1991} and 
\citet{massey} have already been described. In their seminal paper, 
\citet{klaus_spd} introduced the \spd\ method and used V453\,Cyg for its 
very first application. This binary is totally eclipsing, and a spectrum 
taken during totality in the secondary eclipse matched their disentangled 
spectrum of the primary star very well, proving the veracity of \spd. All 
three studies were based on digital spectra, but the analysis methods differed. 
The resulting masses were $M_{\rm A} = 13.0$--14.4\Msun\ and $M_{\rm B} = 
10.6$--11.1\Msun; our new values are $M_{\rm A} = 13.9 \pm 0.2$\Msun\ and 
$M_{\rm B} = 11.0 \pm 0.2$\Msun.

%%%%%%%%%%%%%%%%%%%%%%%%%%%%%%%%%%%%%%%%%%%%%%%%%%%%%%%%%%%%%%%%%%%%%%

\section{Spectroscopic analysis}

\subsection{Reconstruction of the individual components' spectra}

Disentangled spectra of the components are normally still in the common 
continuum of the binary system. This is because the lack of significant 
light variations outside eclipse means that the determinant of the set of 
equations for zero mode in Fourier expansion is strictly zero. This 
singularity corresponds to the intrinsic uncertainty of how to distribute 
the observed continuum flux over the two components given that \spd\ is only 
sensitive to \emph{changes} in the flux distribution of a spectrum. This, 
and near-singularities which exist for other low modes, are responsible for 
undulations often seen in disentangled spectra \citep{hensberge_prague, 
hensberge_2008}. These spurious patterns were first found in \spd\ of 
relatively broad spectral segments (\citealt{hensberge_2000, fitzpatrick}, PH05).

In the case of significant light variability, there is a guarantee that no 
singularities occur. This means that spectra obtained during eclipse help 
to stabilise the solution of \spd. Caution is needed at this point. In the 
most cases, the line profiles during the eclipses could be affected by the 
Rossiter-McLaughlin effect. Distortion of the line profiles violates 
a principal assumption of \spd\ that the components' spectra are not 
variable except for a scaling due to the light variability. Disentangling 
of spectra affected by pulsations, the Rossiter-McLaughlin effect, or any 
other distortions, would produce unreliable results. However, the spectra 
obtained in the total eclipse are extremely useful because then \spd\ has 
access to information on the continuum flux of the stars so is able to 
return disentangled spectra in their individual continua \citep{pav_v453}.

We performed \spd\ in pure separation mode \citep{pav_brno}, with no significant
 light variability even outside the eclipses. Renormalisation of the separated 
spectra from a common continuum of the binary system to the components' spectra 
in their individual continua is a two-step procedure as was elaborated in PH05. 
A linear transformation involves both an additive and multiplicative operation. 
The additive terms come from differing line blocking between the components. 
The multiplication factors could be fixed from the light ratio derived in the 
light curve solution (\citealt{hensberge_2000}, PH05), or determined from the 
optimal fitting of the separated spectra \citep{tamajo_genfit}. In the present 
paper we use both approaches. Obviously, in renormalisation random and systematic 
errors are multiplied by a factor inversely proportional to the light dilution 
factor. For random errors this amplification is counteracted with a gain in S/N 
coming from the use of time-series of the observed spectra. In other words, even 
when the spectral lines in the observed composite spectrum of binary system are 
resolved (e.g.\ at quadrature phases), it is more reliable to fit disentangled 
spectra than a single composite spectrum because of the gain in S/N achieved in 
\spd.

\subsection{Optimal fitting of disentangled spectra}

Model atmospheres are defined by the \Teff, \logg\ and metallicity. Once 
disentangled spectra are transformed from a common continuum level to their 
intrinsic continuum flux (Section 5.2) these stellar parameters could be derived 
with standard spectroscopic tools as for single stars. The atmospheric diagnostics
 include also measurements of $v\sin i$, which is one of the most prominent 
broadening mechanisms for metallic lines. Stark-broadened H lines are not 
affected by the rotational kernel.

For a hot, high-mass star the helium and silicon lines, especially if present 
in different ionisation stages, are sensitive indicators of \Teff. The Balmer 
lines can also serve this purpose, but are often affected with inaccuracy in the 
normalisation of \'echelle spectra since broad wings of these lines can be spread
 over two to three \'echelle orders. The uncertainties of the continuum placement 
and \'echelle order merging are, with the singularity of Fourier zero mode, the 
main source of spurious patterns in disentangled spectra. In short-period binary 
systems, metallic lines are often broadened by a high $v\sin i$ leaving the 
\ion{He}{i} and \ion{He}{ii} lines the best diagnostic lines.

A grid search is the most common practice for determing the optimal stellar 
atmospheric parameters. Here we did the same, but expanded it for an additional 
option. Disentangling spectra preserves the astrophysical contents of the 
individual components. There is a scaling factor between the disentangled 
spectrum and intrinsic spectrum of the component which is equal to the 
component's fractional light contribution to the total light of the system. 
Fortunately, the FWHM as a measure of $v\sin i$ is invariant on this scaling 
factor. Hence, the light ratio between the components could be determined 
from disentangled spectra directly, without a need for external information. 
This is useful in cases when the light ratio is poorly defined in the light 
curve solution (partial eclipses), or the photometric observations have low 
precision. Determination of the light ratio in constrained mode 
\citep{tamajo_genfit}, i.e.\ with the condition that the sum of the light 
dilution factors should be equal to 1, is a fine cross-check for the credibility 
of other methods (e.g.\ the light curve solution).

Our first release of constrained optimal fitting of the disentangled components' 
spectra was the code {\sc genfit} \citep{tamajo_genfit}. For its optimisation 
routine the code uses a genetic algorithm inspired by the {\sc pikaia} subroutine 
of \citet{charbonneau_pikaia}. The error propagation is calculated with the 
Levenberg-Marquart method \citep{press1992}. {\sc starfit} \citep{kolbas_uher} 
is our second release and an expanded version of {\sc genfit}. It uses the same 
optimisation routine, but error calculations are performed by an MCMC routine 
\citep{Ivezic}. With {\sc starfit} the following parameters for each component 
can be either optimised or fixed: \Teff, \logg, $v\sin i$, light dilution factor 
$ldf$, Doppler shift relative to the wavelength rest frame, and an additive 
constant for continuum level adjustment. Both codes can be run in constrained 
mode (simultaneous fit for both components with the condition that $ldf_1 + ldf_2 
= 1.0$) and unconstrained mode (an independent run for each component 
disentangled spectrum). A grid search was made in precalculated spectra (see 
below). Also, we used synthetic NLTE spectra of O and B-type stars calculated 
by \citet{tlusty_ostars, tlusty_bstars}.

\subsection{Atmospheric parameters} \label{sec:atmpar}

\begin{figure}
\centering
\begin{tabular}{cc}
\includegraphics[width=37mm]{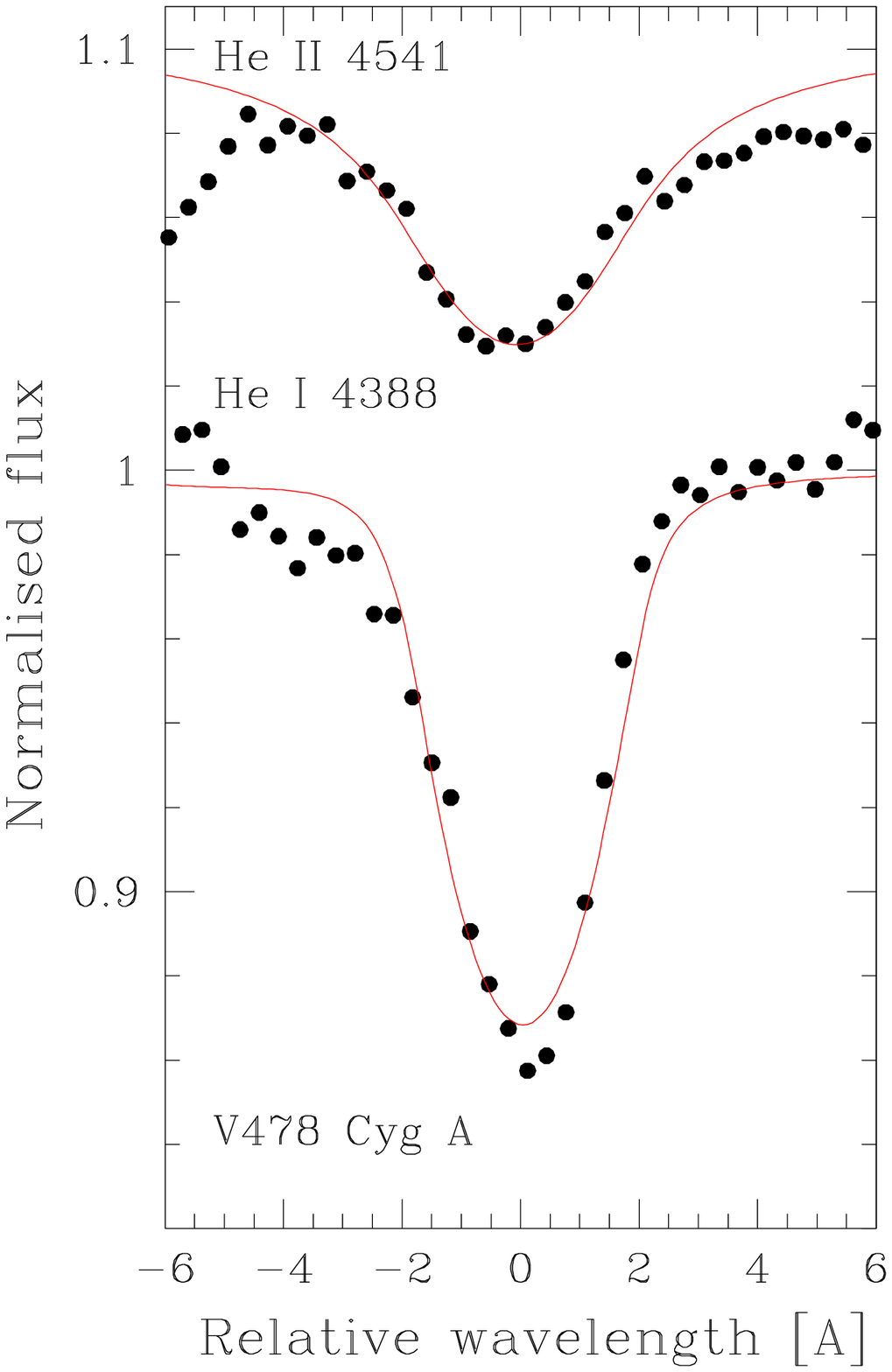} &
\includegraphics[width=37mm]{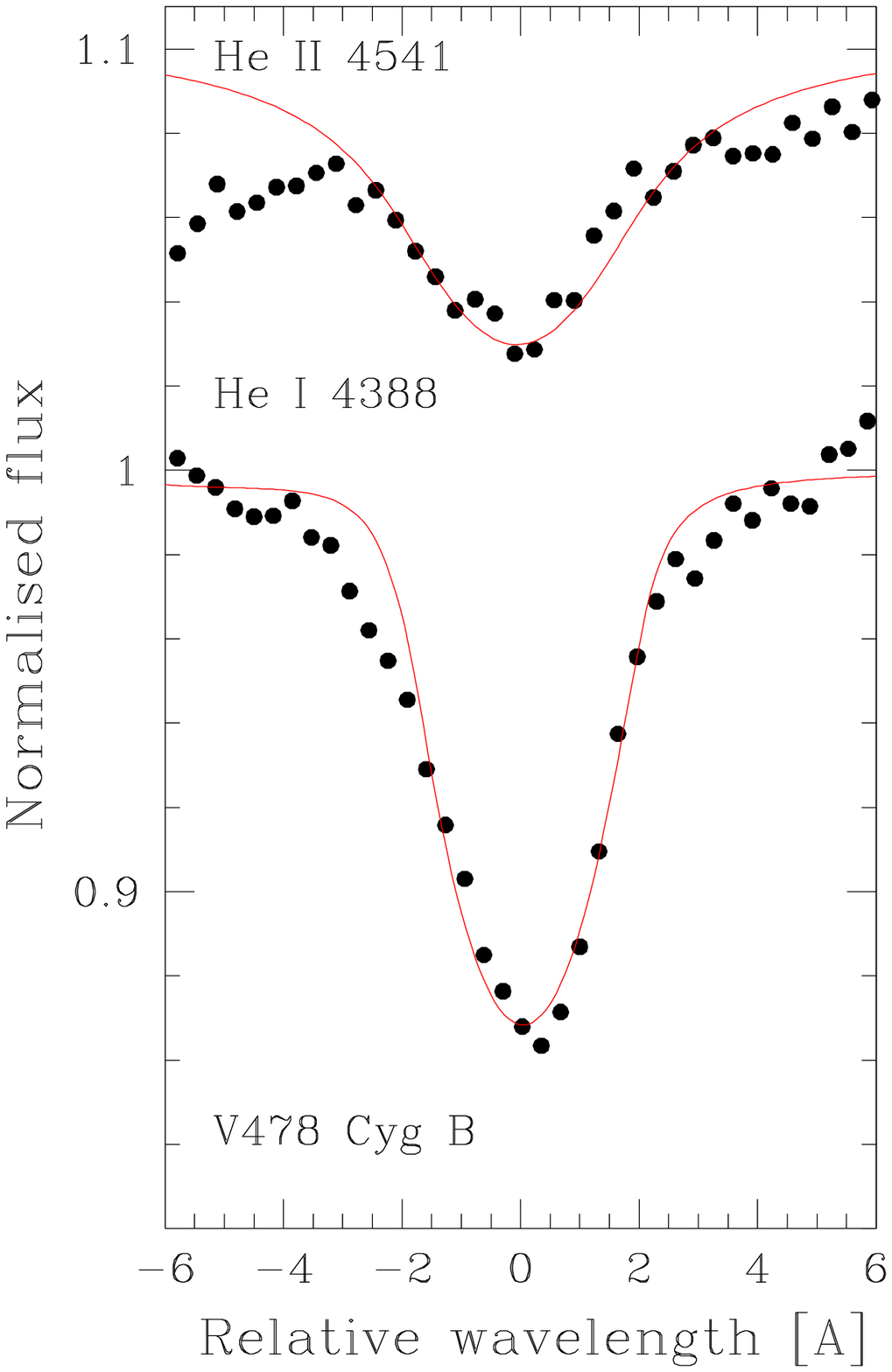} \\
\includegraphics[width=37mm]{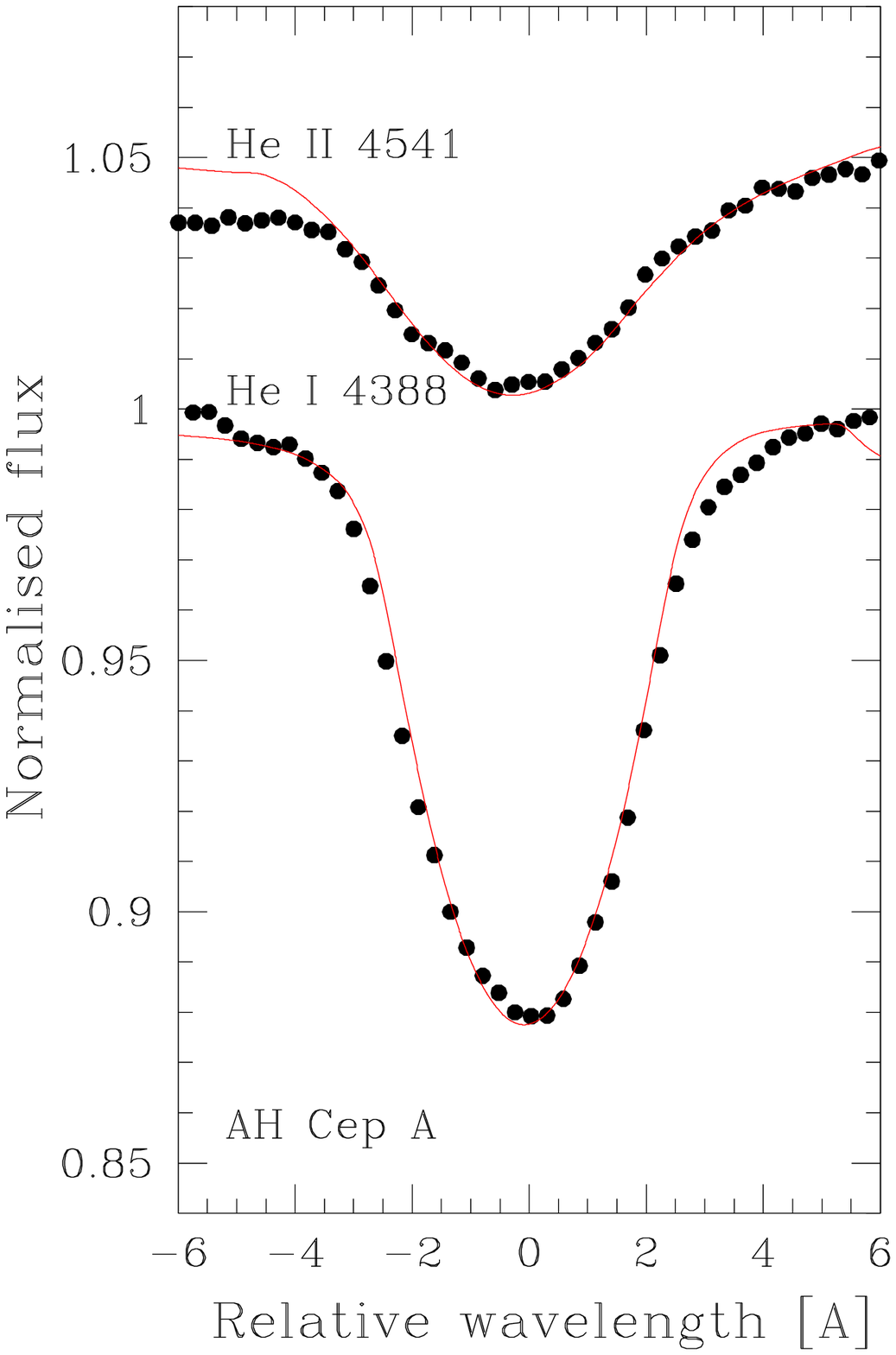} &
\includegraphics[width=37mm]{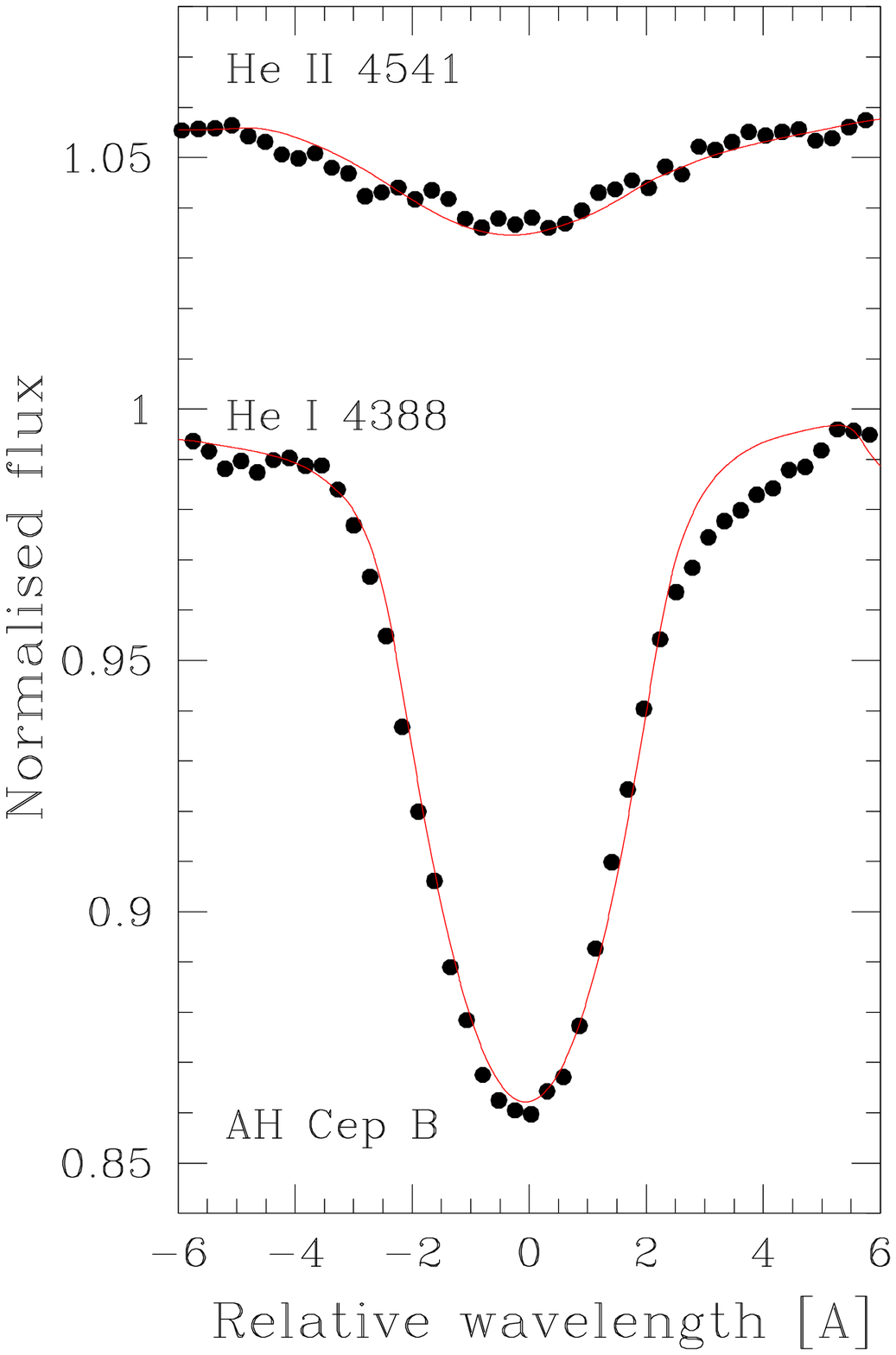} \\
\includegraphics[width=37mm]{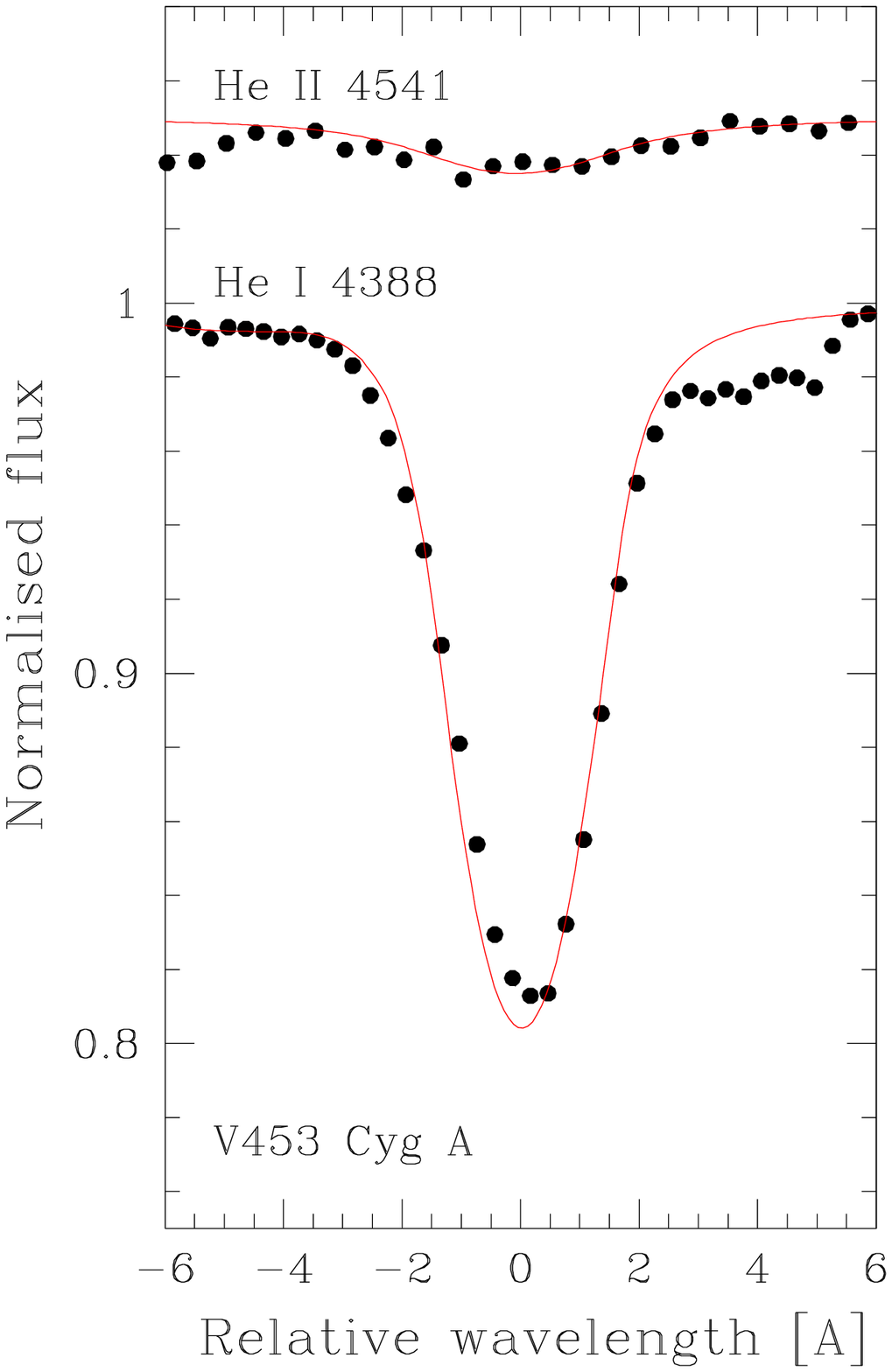} &
\includegraphics[width=37mm]{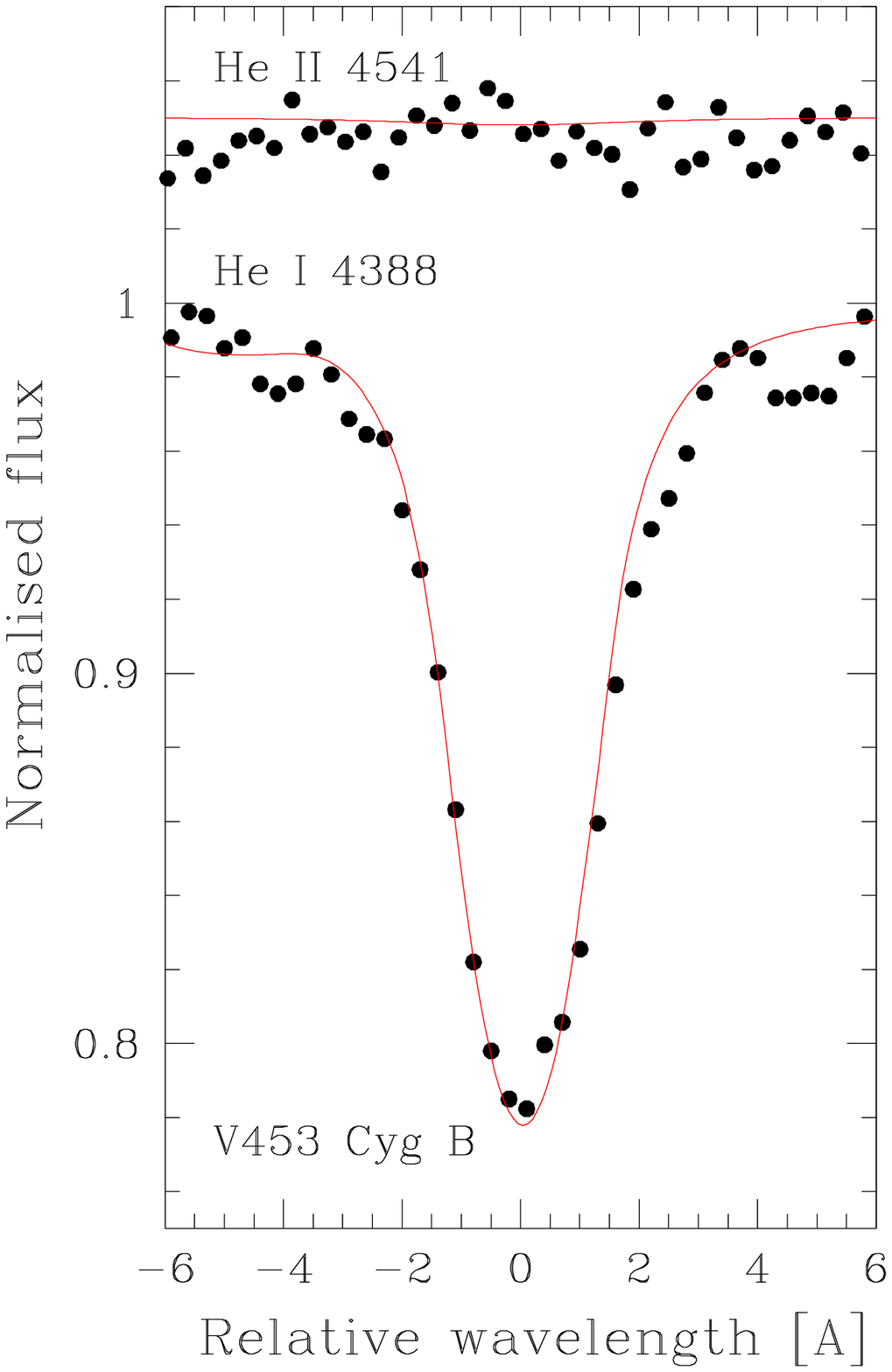} \\
\end{tabular}
\caption{\label{fig:plothel}
Determination of the \Teff s for the components of binary stars studied in the 
present paper. The quality of fits for only the \ion{He}{i} line at 4388\,\AA\ 
and the \ion{He}{ii} line at 4341\,\AA\ are shown. Disentangled spectra are 
represented by the dots, and thoretical line profiles which are the best fit 
to a whole set of the helium lines are shown with red lines.}
\end{figure}

The atmospheric parameters (\Teff,\logg\ and metallicity) are the principal 
ingredients for the generation of the model atmospheres needed for a detailed 
abundance analysis. Our approach has been outlined above, and includes: 
(i) fixing the \logg\ to the value found from the combined spectroscopic 
and light curve analysis; (ii) whenever possible using the light ratio from 
the light curve analysis for the renormalisation of disentangled spectra; 
(iii) determining the $v\sin i$ values from clean (unblended) metal lines. 
Obviously, an iteration procedure is needed to converge for the atmospheric 
parameters. The {\sc starfit} code was used for determination of the atmospheric 
parameters, using H and He lines, and assigning more weight to \ion{He}{i} and 
\ion{He}{ii} lines as they are a sensitive probe of \Teff. Examples of the 
quality of the fits to He lines are shown in Fig.\,\ref{fig:plothel}. The 
results for \Teff\ and $v\sin i$ are given in Table\,\ref{tab:abs_para}.

For a check of the consistency of the spectroscopically derived quantities 
with those from the iterative procedure, we used {\sc starfit} in both 
constrained and unconstrained modes. Fixing \logg\ was found to be especially 
useful due to the degeneracy between \Teff\ and \logg\ when fitting Balmer 
line profiles in hot stars. It was encouraging to find a good agreement for 
the light ratio from disentangled spectra and the light curve solutions. 
Therefore, in the light curve analysis of AH\,Cep we opted for the 
spectroscopically determined light ratio (Section~\ref{sec:lc:ah}).

\begin{table*}
\centering
\caption{\label{tab:abs_para}
Astrophysical quantities derived in this work, expect for V578\,Mon where the 
values from \citet{Garcia_2014} are given for completeness. The listed parameters
 are determined in complementary analysis of the spectra and light curves.}
\begin{tabular}{lcccccccc} \hline
Star   &  $M$  & $R$  & $\log g$  &  $T_{\rm eff}$  & $\log L$ & $v\,\sin\,i$ & $v_{\rm synch}$ & $\xi_{\rm t}$ \\
       &\Msun\ & R$_{\sun}$  & [cgs]  & [K]       & L$_{\sun}$ & [km\,s$^{-1}$] &  [km\,s$^{-1}$] &  [km\,s$^{-1}$] \\
\hline
V478\,Cyg A & 15.40 $\pm$ 0.38 & 7.26 $\pm$ 0.09 & 3.904 $\pm$ 0.009 & 32\,100 $\pm$ 550 & 4.70 $\pm$ 0.03 & 129.1 $\pm$ 3.6 & 127.4 $\pm$ 1.5 &  5 $\pm$ 1 \\
V478\,Cyg B & 15.02 $\pm$ 0.35 & 7.15 $\pm$ 0.09 & 3.907 $\pm$ 0.010 & 31\,800 $\pm$ 600 & 4.67 $\pm$ 0.04 & 127.0 $\pm$ 3.5 & 125.5 $\pm$ 1.6 &  5 $\pm$ 1 \\
AH\,Cep A   & 16.14 $\pm$ 0.26 & 6.51 $\pm$ 0.10 & 4.019 $\pm$ 0.012 & 30\,700 $\pm$ 550 & 4.53 $\pm$ 0.03 & 172.1 $\pm$ 2.1 & 185.4 $\pm$ 1.0 &  6 $\pm$ 1 \\
AH\,Cep B   & 13.69 $\pm$ 0.21 & 5.64 $\pm$ 0.11 & 4.073 $\pm$ 0.018 & 28\,800 $\pm$ 630 & 4.30 $\pm$ 0.04 & 160.6 $\pm$ 2.3 & 160.6 $\pm$ 1.0 &  3 $\pm$ 1 \\
V453\,Cyg A & 13.90 $\pm$ 0.23 & 8.62 $\pm$ 0.09 & 3.710 $\pm$ 0.009 & 28\,800 $\pm$ 500 & 4.66 $\pm$ 0.04 & 107.2 $\pm$ 2.8 & 112.1 $\pm$ 1.2 & 14 $\pm$ 1 \\
V453\,Cyg B & 11.06 $\pm$ 0.18 & 5.45 $\pm$ 0.08 & 4.010 $\pm$ 0.012 & 27\,700 $\pm$ 600 & 4.20 $\pm$ 0.05 &  98.3 $\pm$ 3.7 &  70.8 $\pm$ 1.0 &  3 $\pm$ 1 \\
V578\,Mon A & 14.54 $\pm$ 0.08 & 5.41 $\pm$ 0.04 & 4.133 $\pm$ 0.018 & 30\,000 $\pm$ 500 & 4.33 $\pm$ 0.03 & 123   $\pm$ 5   & 110.0 $\pm$ 3.1 &  4 $\pm$ 1 \\
V578\,Mon B & 10.29 $\pm$ 0.06 & 4.29 $\pm$ 0.05 & 4.185 $\pm$ 0.021 & 25\,750 $\pm$ 435 & 3.86 $\pm$ 0.03 &  99   $\pm$ 3   &  89.1 $\pm$ 2.1 &  2 $\pm$ 1 \\
\hline \\
\end{tabular}
\end{table*}

\begin{table*}
\centering
\caption{\label{tab:abund}
Abundances determined for the stars in our sample of binary systems. The atmospheric parameters
used for the construction of model atmospheres are given in Table\,\ref{tab:abs_para}.}
\begin{tabular}{lccccccc} \hline
Star   &  C  & N  & O  & [N/C]  &  [N/O] & Mg & Si  \\
\hline
V478\,Cyg A  & 8.24 $\pm$ 0.09 & 7.69 $\pm$ 0.11 & 8.65 $\pm$ 0.08 & $-1.03 \pm 0.17$ & $-0.55 \pm 0.15$ & 7.69 $\pm$ 0.10 & 7.58 $\pm$ 0.11 \\
V478\,Cyg B  & 8.28 $\pm$ 0.08 & 7.68 $\pm$ 0.09 & 8.72 $\pm$ 0.11 & $-0.99 \pm 0.12$ & $-0.60 \pm 0.12$ & 7.70 $\pm$ 0.10 & 7.59 $\pm$ 0.06 \\
AH\,Cep A    & 8.37 $\pm$ 0.08 & 7.64 $\pm$ 0.07 & 8.66 $\pm$ 0.12 & $-1.02 \pm 0.14$ & $-0.73 \pm 0.11$ & 7.78 $\pm$ 0.10 & 7.69 $\pm$ 0.11 \\
AH\,Cep B    & 8.27 $\pm$ 0.05 & 7.67 $\pm$ 0.06 & 8.69 $\pm$ 0.12 & $-1.02 \pm 0.13$ & $-0.60 \pm 0.08$ & 7.52 $\pm$ 0.10 & 7.47 $\pm$ 0.09 \\
V453\,Cyg B  & 8.28 $\pm$ 0.04 & 7.73 $\pm$ 0.09 & 8.74 $\pm$ 0.09 & $-1.01 \pm 0.13$ & $-0.55 \pm 0.10$ & 7.53 $\pm$ 0.10 & 7.66 $\pm$ 0.06 \\
V453\,Cyg B  & 8.24 $\pm$ 0.06 & 7.76 $\pm$ 0.08 & 8.74 $\pm$ 0.08 & $-0.98 \pm 0.13$ & $-0.48 \pm 0.10$ & 7.54 $\pm$ 0.10 & 7.54 $\pm$ 0.08 \\
V578\,Mon A  & 8.18 $\pm$ 0.07 & 7.69 $\pm$ 0.12 & 8.74 $\pm$ 0.10 & $-1.05 \pm 0.16$ & $-0.49 \pm 0.14$ & 7.52 $\pm$ 0.10 & 7.50 $\pm$ 0.06 \\
V578\,Mon B  & 8.21 $\pm$ 0.11 & 7.72 $\pm$ 0.09 & 8.76 $\pm$ 0.11 & $-1.04 \pm 0.14$ & $-0.49 \pm 0.14$ & 7.50 $\pm$ 0.10 & 7.44 $\pm$ 0.06 \\
\hline
OB binaries  & 8.26 $\pm$ 0.05 & 7.70 $\pm$ 0.04 & 8.71 $\pm$ 0.04 & $-1.01 \pm 0.06$ & $-0.56 \pm 0.06$ & 7.59 $\pm$ 0.08 & 7.57 $\pm$ 0.10 \\
B stars$^1$  & 8.33 $\pm$ 0.04 & 7.79 $\pm$ 0.04 & 8.76 $\pm$ 0.05 & $-0.97 \pm 0.06$ & $-0.54 \pm 0.06$ & 7.56 $\pm$ 0.05 & 7.50 $\pm$ 0.05 \\
\hline \\
\end{tabular}
\end{table*}

\subsection{Photospheric chemical composition}

We have estimated the chemical abundances of the stars by fitting the 
renormalized disentangled spectra with synthetic spectra. Non-LTE line 
formation and spectrum synthesis computations were performed using {\sc detail} 
and {\sc surface} \citep{giddings,butler} and the model atoms as listed in 
\citet{pav_v453}. Model atmospheres were calculated with {\sc atlas9} 
\citep{kurucz1979}. Justification for this hybrid approach can be found in 
\citet{fernanda_hybrid}.

\subsubsection{V478\,Cyg}

V478\,Cyg contains stars with very similar fundamental quantities. The 
differences in mass and \Teff\ are only 0.4\Msun\ and 600\,K, respectively, 
within their 1$\sigma$ uncertainties. The abundances derived for the stars 
are also identical to within 1$\sigma$. The abundance pattern is close to the 
`present-day cosmic' abundances \citep{fernanda_standard}, except for magnesium. 
The Mg abundances are $\log \epsilon({\rm Mg}) = 7.69 \pm 0.10$ for the primary 
and $\log \epsilon({\rm Mg}) = 7.70 \pm 0.10$ for the secondary, so are 
internally very consistent. They are somewhat larger than the `standard' Mg  
abundance in slowly rotating B-stars in the solar neighbourhood, $\log 
\epsilon({\rm Mg}) = 7.56 \pm 0.05$ \citep{fernanda_standard}. The Mg 
abundance in V478\,Cyg is almost identical to the mean Mg abundance determined 
for 52 B-stars by \citet{lyubimkov_mag}. However, they adopted the mean abundance 
of $\log \epsilon({\rm Mg}) = 7.59 \pm 0.15$ from the stars with the most 
reliably determined microturbulence velocities ($\xi_{\rm t}$). Using the Mg 
abundance as a good indicator of metallicity \citep{lyubimkov_mag}, we can 
conclude that the metallicity of the stars in V478\,Cyg is slightly above solar, 
[Mg/H] $= 0.15 \pm 0.10$, using $\log \epsilon {\rm (Mg)}_{\odot}  
= 7.55 \pm 0.02$ from \citet{asplund}.

The CNO abundances for the components of V478\,Cyg were determined recently 
by \citet{martins_mahy}, who found considerable differences of 0.30\,dex (C), 
0.36\,dex (N) and 0.12\,dex (O) which are not corroborated by our own 
determinations (Table\,{\ref{tab:abund}). Our log(N/O) and log(N/C) also 
clearly disagree with the results of \citet{martins_mahy}. There is a 
difference between the mass ratios derived in the two studies, and 
\citet{martins_mahy} found a larger mass difference of $\sim$0.7\Msun\ 
than our own. \citet{martins_mahy} found the secondary star to be more 
luminous and also about 1000\,K cooler than the primary, which is difficult 
to understand for a MS binary system. The abundance measurements are sensitive 
to the $\xi_{\rm t}$ used: this is not reported by \citet{martins_mahy} so we 
do not make any further comparison with this study.

\subsubsection{AH\,Cep}

The measured abundances for the components of AH\,Cep are consistent for CNO, 
but not for Mg and Si. Whilst the secondary star is of solar metallicity, if 
the Mg abundance is used as a metallicity indicator, the primary star is 
supersolar, with [Mg/H] $=$ 0.36 $\pm$ 0.10 dex. Also, the primary's Si 
abundance is higher than solar \citep{asplund} and the 'present-day cosmic' 
standard abundance \citep{fernanda_standard}. Under the assumption of the 
same initial chemical composition for the components of binary system, it is 
hard to explain how stars of such young age differ in metallicity. An incorrect 
light ratio, or dilution by some third light could explain these differences, 
but any corrections for these effects would also affect determination of the 
atmospheric parameters, and thus the CNO abundances. In a comprehensive study 
of the Mg abundance in B stars \citet{lyubimkov_mag} examined the influnce of 
the \ion{Al}{iii} line at 4480.1\,\AA, which blends with the usually much 
stronger \ion{Mg}{ii} 4481.2\,\AA\ line. For the \Teff s of the components 
in AH\,Cep this is unlikely to cause the discrepancies.

\citet{martins_mahy} also studied AH\,Cep, and determined the CNO abundances 
but unfortunately not those for Mg and Si. On the average the C abundances 
for the components between both studies agree, but the N and O abundances 
from \citet{martins_mahy} are systematically lower. Compared to our results 
(Table~\ref{tab:abund}) they obtained an underabundance for both species, 
with $\Delta \log \epsilon({\rm N}) = -0.34$ and $-0.13$ dex, and $\Delta 
\log \epsilon({\rm O}) = -0.21$ and $-0.33$ dex, for the primary and secondary 
star, respectively. Again, without knowledge of the $\xi_{\rm t}$s used by 
\citet{martins_mahy} it is hard to trace the deviations between the two studies.

\subsubsection{V578\,Mon}

Th {\sc hermes} spectra used in this work cover almost the entire optical 
spectral region, and substantially extend the spectral coverage on which 
the previous spectroscopic analysis by PH05 was based. Since the {\sc caspec}
 spectra used by PH05 cover only 4000--5000\,\AA\ the new data makes important 
groups of N and C lines accessible in the 4990--5050\,\AA\ and 5115--5150\,\AA\ 
regions, plus some lines of interest in other spectral regions. Since the $v\sin 
i$ values for the components are not as large (100--120\kms), line blending is 
less severe for V578\,Mon than in the other objects studied in the current work. 
This is important since C and N lines in this \Teff\ range are quite weak; the 
measured abundances for V578\,Mon are more secure.

The results of the abundance analysis for the components of V578\,Mon are given 
in Table~\ref{tab:abund}. The photospheric abundance pattern of the two components
 is almost identical, and for all studied species is identical to within 
1$\sigma$. Both components share the `cosmic abundance standard' 
\citep{fernanda_standard}, which is given at the base of Table~\ref{tab:abund}, 
except for the C abundances which are within 2$\sigma$.

The photospheric composition for the components of V578\,Mon was determined by 
PH05 relative to the sharp-lined B-type star \#201 in the open cluster NGC~2244, 
for which a detailed spectroscopic analysis was performed by \citet{vrancken}. 
Star \#201 has $\Teff = 27\,300 \pm 1000$\,K and $\logg = 4.3 \pm 0.1$ and 
closely resembles V578\,Mon\,B. PH05 rotationally broadened the CASPEC spectrum 
of star \#201 to measured the $v\sin i$ value of the components of V578\,Mon. 
The derived abundances matched the abundance pattern of \#201 fairly well, and 
within the 1$\sigma$ uncertainties. However, our new determinations do not 
corroborate these findings. Beside the C abundances, all other species are 
systematically underabundant. In this context only the deviation in abundance 
for Si could be understandable: in previous studies all available Si lines were 
used, and this is known to produce underestimates due to uncertainties in the 
Si model atom. In this work we used a selection of Si lines after the critical 
examination by \citet{sergio}. For the \Teff s of the stars in question the 
C lines change only weakly, which could explain the apparent consistency between 
the new and previous determinations.

A source of the deviations could be the $\xi_{\rm t}$, but we cannot easily 
trace its systematics since the O abundances for the components of V578\,Mon 
in the previous study are in perfect agreement with those in star \#201. In 
the present study we use the \Teff s of the components found by 
\citet{Garcia_2014}, which are basically the same as those originally 
determined in \citet{hensberge_2000} and used by PH05. One avenue for resolving 
the situation is to reanalyse the reference star \#201 based on new high-quality 
\'echelle spectra. Unfortunately, possible systematic errors in the abundances 
for star \#201 make obsolete any further comparison of the abundance pattern 
found by \citet{vrancken} for early-B stars in NGC~2244.

\subsubsection{V453\,Cyg}

The previous abundance determination for this object \citep{pav_v453} suffers 
from a limited spectral coverage of only 4000--4900\,\AA. As a consequence the 
C abundance relied on only two spectral lines, which were also blended with 
O lines. As with V578\,Mon, it it important to extend the wavelength coverage 
to beyond H$\beta$.

For V453\,Cyg\,A, the most evolved star in the current work but still on the MS, 
\citet{pav_v453} found a very high $\xi_{\rm t}$ of $15 \pm 1$\kms. So far, only 
V380\,Cyg\,A (an evolved star on the subgiant branch; \citealt{andrew_v380}), 
has been found with such a high $\xi_{\rm t}$. It is encouraging that our 
reanalysis of V453\,Cyg\,A yields $\xi_{\rm t} = 14 \pm 1$\kms, which confirms 
the previous finding. Also, it should be noted that the \Teff\ measurement for 
the primary star has also been revised to better match the \ion{He}{ii} lines, 
and is now $28\,700 \pm 550$\,K. The C abundance is now much more reliably 
determined, primarily from lines in the 5130--5160\,\AA\ region. We also 
performed a check with disentangled profiles of spectral segments centred 
on H$\alpha$, and the \ion{C}{ii} lines at 6555\,\AA\ and 6568\,\AA. Since 
we avoid a weak \ion{Si}{iii} triplet on a blue wing of the H$\beta$ line, 
the Si abundances for both components are now consistent with their Mg 
abundances (Table~\ref{tab:abund}).

%%%%%%%%%%%%%%%%%%%%%%%%%%%%%%%%%%%%%%%%%%%%%%%%%%%%%%%%%%%%%%%%%%%%%%

\section{Light curve analysis}

The light curves of V478\,Cyg, AH\,Cep and V453\,Cyg were modelled using the 
Wilson-Devinney (WD) code \citep{Wilson1971,Wilson1979} in order to determine 
their photometric parameters. This code implements full Roche geometry and 
accounts for the effects of reflection, gravitational distortion, and limb 
and gravity darkening. We used the {\sc jktwd} wrapper \citep{jkt_1066}, 
which implements automatic iteration to the differential-corrections procedure 
in the 2004 version of the WD code (hereafter called {\sc wd2004}).

In order to determine the best solution for each light curve, we defined 
a default solution based on a typical approach to modelling dEBs. We then 
systematically varied each of the input parameters and available control 
parameters for {\sc wd2004} and reran the least-squares fit, to find those 
that had a significant effect on the output parameters. The errorbar for 
each output parameter was obtained from the spread of results for that 
parameter coming from the different fits. These errorbars are almost always 
much larger than the formal errors calculated by {\sc wd2004} from the 
covariance matrix, underlining the inadequacy of the formal errors in 
capturing the true uncertainty of the parameter values. Informed by these 
investigations, we then calculated final photometric solutions which represent 
our best physical interpretation of the light curves of each system.

For our default solution we chose Mode 0 in {\sc wd2004} 
\citep[see][]{Wilson2004}, which is for a detached binary where the relative 
light contributions of the two stars to each of the light curves are not 
coupled to each other or to the \Teff\ values. We left the \Teff s fixed at 
the values found in Section~\ref{sec:atmpar}. The mass ratios, eccentricities 
and arguments of periastron were fixed to the spectroscopic values. All light 
curves were fitted together. The rotation rates were fixed to the values found 
from the atmospheric analysis. Bolometric albedos and gravity brightening 
exponents were fixed to 1.0 as expected for radiative envelopes. A linear 
limb darkening (LD) law was used, and coefficients were fixed at values 
interpolated from the tables of \citet{Vanhamme}. Our exploratory solutions 
used a reasonable numerical precision ({\sc n1} $=$ {\sc n2} $=$ 30; 
see \citealt{Wilson2004}).

To obtain the final `best' solutions we used a high numerical precision 
({\sc n1} $=$ {\sc n2} $=$ 60), the simple model of reflection, and suitable 
orbital ephemerides. We fitted for the potentials of the two stars, their 
contributions to the total light in each passband, orbital inclination, and 
the phase shift of the primary eclipse.

For the exploration and error-analysis phase we performed each of the following 
checks. (1) Solutions were obtained in Mode 2, which couples the \Teff s to the
 light contributions in each passband using predictions from model atmospheres. 
(2) Each light curve was modelled individually. (3) We varied the mass ratio by 
the size of its error bar from the spectroscopic analysis. (4) We varied the 
eccentricity and argument of periastron by the size of their error bars from 
the spectroscopic analysis. (5) Solutions were obtained using the detailed 
reflection option. (6) The rotation rates were included as fitted parameters. 
(7) The albedos were included as fitted parameters. (8) The gravity brightening 
exponents were included as fitted parameters. (9) The linear LD coefficients 
were fitted. (10) The logarithmic and square-root LD laws were tried. (11) 
Third light was included as a fitted parameter.

\subsection{V478\,Cyg}

This system shows apsidal motion so the argument of periastron was included
 as a fitted parameter in all solutions. The orbital eccentricity was fixed 
at 0.021 based on our spectroscopic solution.

Light curves are available in the $B$ and $V$ bands from \citet{sezer}, 
totalling 1094 and 1079 datapoints respectively, and it is these data on 
which our solution is based. We were not able to obtain the data from 
\citet{popper_dumont}. A $V$-band light curve containing 418 points was 
obtained by \citet{Tartu} using a 0.25\,m telescope in New Mexico, in the 
framework of a variability survey of OB associations. The measurements are 
quite scattered and also sparsely sampled so were not considered in detail. 
Finally, $U$, $B$, $V$ and $R$ light curves of V478\,Cyg were obtained by 
\citet{zakirov_v478}, totalling 712, 719, 717 and 718 datapoints. These 
have full phase coverage but a large scatter so were not used in our best 
solution.

The \citet{sezer} $BV$ light curves do not yield a determinate ratio of the 
radii because the eclipses are partial and quite shallow. We therefore obtained 
a light ratio between the two stars from the spectroscopic analysis and used 
this to break the solution degeneracy. The uncertainty in this light ratio was 
propagated into the final results by running solutions with the light ratio 
perturbed by its errorbar. We expected that third light would be significant 
for this system, but the best-fitting value of third light is small and 
consistent with zero. Solutions for the $B$, $V$ and $B+V$ datasets were 
very consistent.

We found that changing the stellar rotational velocities in the {\sc wd2004} 
fit can have a significant effect on the best-fit parameter values. The 
spectroscopic results indicate that the stars rotate super-synchronously. 
We set the rotation rates at 1.40 and 1.33 times the synchronous values and 
determined the effect of rotation on the solution by perturbing these values 
by 5\%, a conservatively large amount.

Whilst the gravity brightening exponents have little effect on the solution, 
the best-fitting values for both stars are 0.85 rather than the 1.0 expected 
on theoretical grounds. The formal error on the values is $\pm$0.2 for both 
stars, so this deviation is not significant. We found that different treatments 
of reflection and LD have a negligible effect on the results. Fitting for the 
LD coefficients does give a slightly better fit to the data, so we included 
them as fitted parameters in the final solution. The results are given in 
Table\,\ref{tab:wd} and the best fits are shown in Fig.\,\ref{fig:lc:v478}.

\begin{table*} \label{tab:wd}
\caption{\label{tab:wd}Summary of the parameters for the {\sc wd2004} solutions of the 
light curves of the systems. Detailed descriptions of the control parameters 
can be found in the {\sc wd2004} user guide \citep{Wilson2004}. A and B refer 
to the primary and secondary stars, respectively. Uncertainties are only quoted 
when they have been robustly assessed by comparison of a full set of alternative 
solutions.}
\begin{tabular}{llccc} \hline
Parameter                           & {\sc wd2004} name     & V478\,Cyg                & AH\,Cep                     & V453\,Cyg                   \\
\hline
{\it Control and fixed parameters:} \\
{\sc wd2004} operation mode         & {\sc mode}            & 0                        & 0                           & 0                           \\
Treatment of reflection             & {\sc mref}            & 1                        & 1                           & 1                           \\
Number of reflections               & {\sc nref}            & 1                        & 1                           & 1                           \\
Limb darkening law                  & {\sc ld}              & 1 (linear)               & 1 (linear)                  & 1 (linear)                  \\
Numerical grid size (normal)        & {\sc n1, n2}          & 60                       & 60                          & 60                          \\
Numerical grid size (coarse)        & {\sc n1l, n2l}        & 30                       & 30                          & 30                          \\[3pt]
{\it Fixed parameters:} \\
Orbital period (d)                  & {\sc period}          & 2.88090063               & 1.774761                    & 3.88982486                  \\
Primary eclipse time (HJD)          & {\sc hjd0}            & 2444777.4852             & 2445962.7359                & 2439340.0993                \\
Mass ratio                          & {\sc rm}              & 0.976                    & 0.848                       & 0.795                       \\
$T_{\rm eff}$ star\,A (K)           & {\sc tavh}            & 32\,100                  & 30\,700                     & 27\,700                     \\
$T_{\rm eff}$ star\,B (K)           & {\sc tavh}            & 31\,800                  & 28\,800                     & 26\,400                     \\
Rotation rates                      & {\sc f1, f2}          & 1.40, 1.33               & 1.0, 1.0                    & 1.59, 2.35                  \\
Gravity darkening                   & {\sc gr1, gr2}        & 1.0, 1.0                 & 1.0, 1.0                    & 1.0, 1.0                    \\
Bolometric albedos                  & {\sc alb1, alb2}      & 1.0, 1.0                 & 1.0, 1.0                    & 1.0, 1.0                    \\
Bolometric LD coeff.\ A             & {\sc xbol1}           & 0.538                    & 0.552                       & 0.585                       \\
Bolometric LD coeff.\ B             & {\sc xbol2}           & 0.542                    & 0.581                       & 0.630                       \\
Orbital eccentricity                & {\sc e}               & 0.021                    & 0.0                         & 0.022                       \\[3pt]
{\it Fitted parameters:} \\
Phase shift                         & {\sc pshift}          & $-$0.0005                & 0.0004                      & 0.0365                      \\
Star\,A potential                   & {\sc phsv}            & 4.714                    & 3.867                       & 4.388                       \\
Star\,B potential                   & {\sc phsv}            & 4.718                    & 4.013                       & 5.629                       \\
Orbital inclination (\degr)         & {\sc xincl}           & $78.23 \pm 0.21$         & $69.81 \pm 0.60$            & $87.51 \pm 0.28$            \\
Longitude of periastron (\degr)     & {\sc perr0}           & $63.4 \pm 4.2$           & --                          & $303 \pm 16$                \\
$U$-band limb darkening coefficient & {\sc ldu}             &                          &                             & 0.369, 0.268                \\
$B$-band limb darkening coefficient & {\sc ldu}             & 0.36, 0.268              &                             & 0.369, 0.268                \\
$V$-band limb darkening coefficient & {\sc ldu}             & 0.23, 0.110              &                             & 0.239, 0.110                \\
$u$-band limb darkening coefficient & {\sc ldu}             &                          & 0.294, 0.302                &                             \\
$v$-band limb darkening coefficient & {\sc ldu}             &                          & 0.279, 0.288                &                             \\
$b$-band limb darkening coefficient & {\sc ldu}             &                          & 0.261, 0.269                &                             \\
$y$-band limb darkening coefficient & {\sc ldu}             &                          & 0.242, 0.236                &                             \\
$U$-band light contributions        & {\sc hlum}, {\sc clum}&                          &                             & 14.17, 5.24                 \\
$B$-band light contributions        & {\sc hlum}, {\sc clum}& 5.615, 5,438             &                             & 13.91, 5.30                 \\
$V$-band light contributions        & {\sc hlum}, {\sc clum}& 7.925, 7.667             &                             & 14.00, 5.29                 \\
$u$-band light contributions        & {\sc hlum}, {\sc clum}&                          & 6.369,2.544                 &                             \\
$v$-band light contributions        & {\sc hlum}, {\sc clum}&                          & 5.815, 2.372                &                             \\
$b$-band light contributions        & {\sc hlum}, {\sc clum}&                          & 5.485, 3.905                &                             \\
$y$-band light contributions        & {\sc hlum}, {\sc clum}&                          & 8.525, 3.467                &                             \\
$U$-band third light                & {\sc el3}             &                          & 0.0                         & 0.019                       \\
$B$-band third light                & {\sc el3}             & 0.0                      & 0.0                         & 0.032                       \\
$V$-band third light                & {\sc el3}             & 0.0                      & 0.0                         & 0.032                       \\
$u$-band third light                & {\sc el3}             &                          & 0.0                         &                             \\
$v$-band third light                & {\sc el3}             &                          & 0.0                         &                             \\
$b$-band third light                & {\sc el3}             &                          & 0.0                         &                             \\
$y$-band third light                & {\sc el3}             &                          & 0.0                         &                             \\
Fractional radius of star\,A        &                       & $0.2728 \pm 0.0024$      & $0.3400 \pm 0.0047$         & $0.2834 \pm 0.0026$         \\
Fractional radius of star\,B        &                       & $0.2686 \pm 0.0028$      & $0.2946 \pm 0.0059$         & $0.1790 \pm 0.0024$         \\[3pt]
\hline
\end{tabular}
\end{table*}

\begin{figure}
\includegraphics[width=\columnwidth]{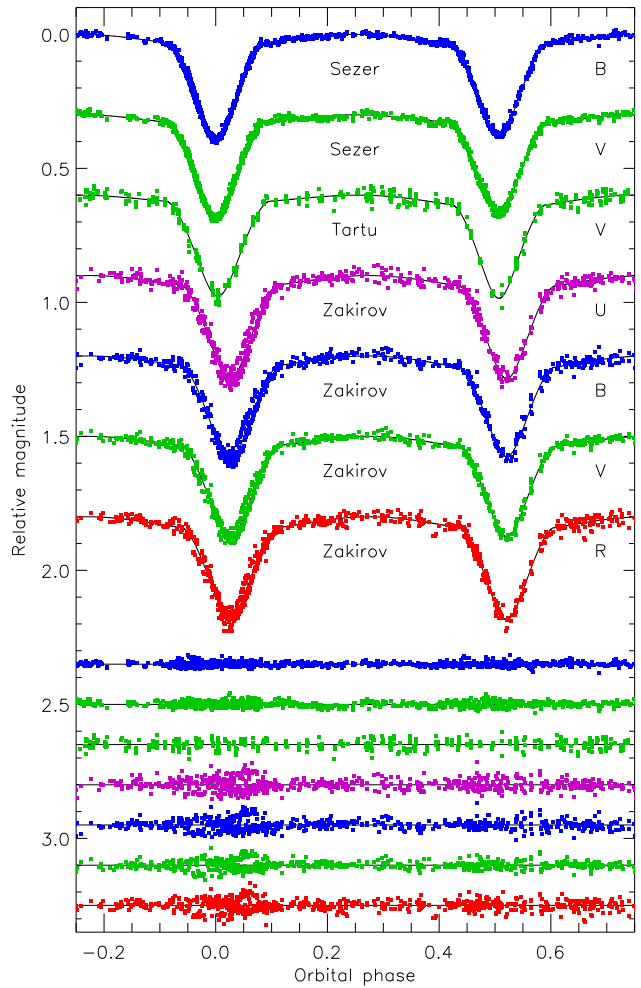} \\
\caption{\label{fig:lc:v478} The light curve solution for V478\,Cyg. The
differential magnitudes are plotted versus orbital phase and colour-coded
according to passband. The source and passband of each light curve is
labelled. The residuals of the fit are shown at the base of the figure.
Offsets in  magnitude have been applied to separate the light curves and
residuals for clarity.}
\end{figure}

\begin{figure}
\includegraphics[width=\columnwidth]{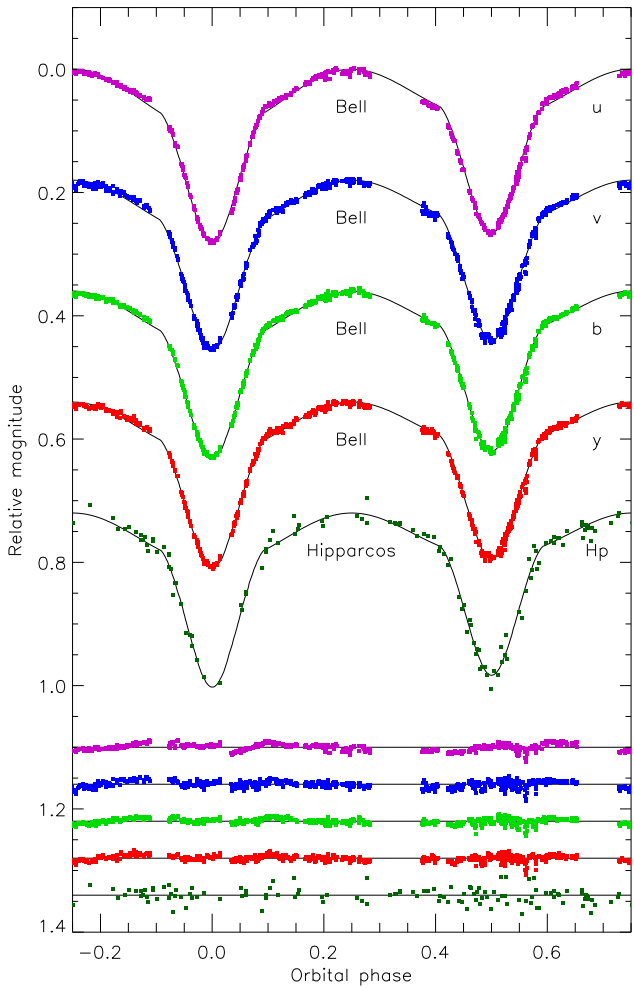} \\
\caption{\label{fig:lc:ah} The light curve solution for AH\,Cep.
Comments are as for Fig.\,\ref{fig:lc:v478}.}
\end{figure}

\begin{figure}
\includegraphics[width=\columnwidth]{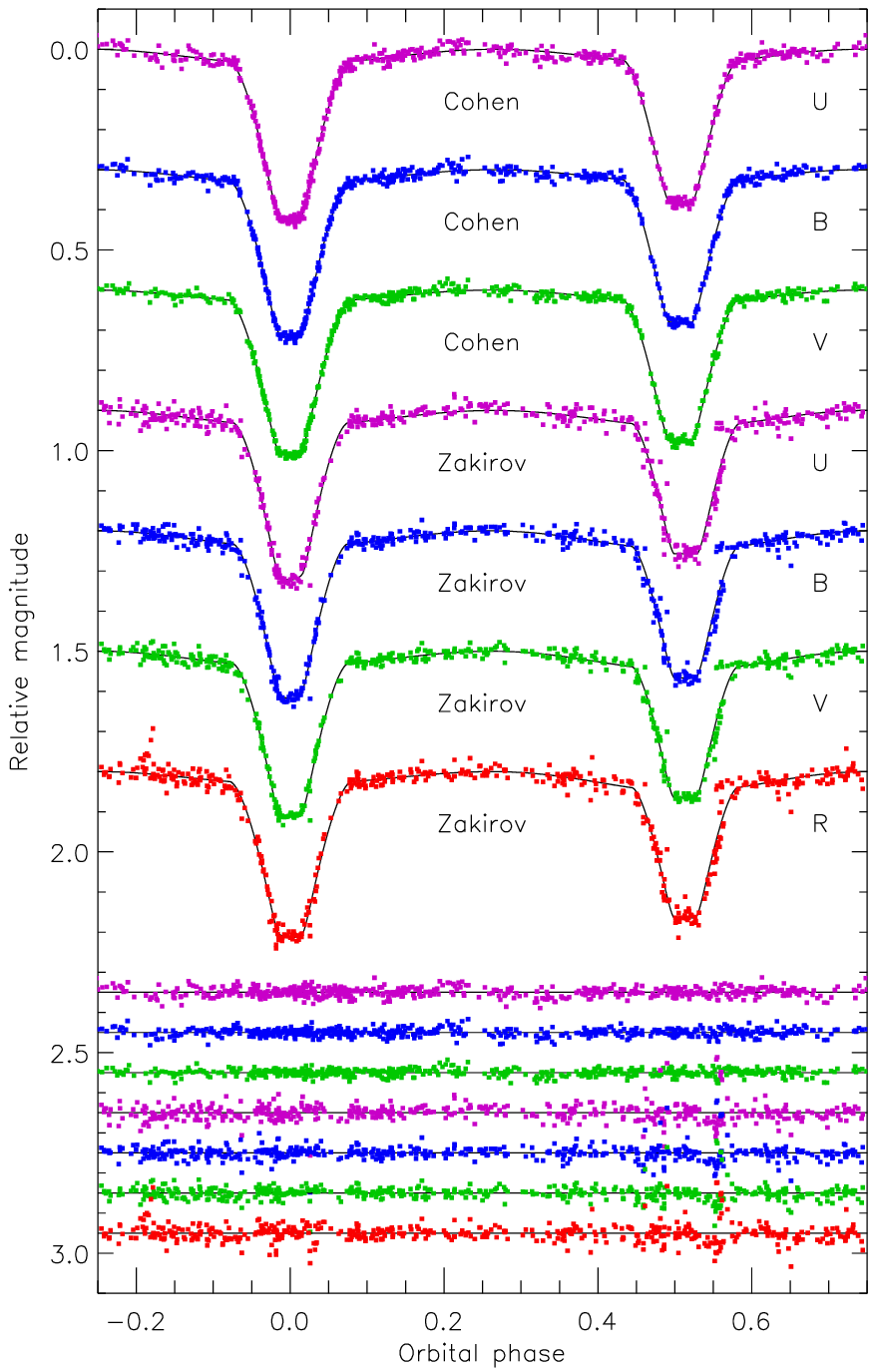} \\
\caption{\label{fig:lc:v453} The light curve solution for V453\,Cyg.
Comments are as for Fig.\,\ref{fig:lc:v478}.}
\end{figure}

\subsection{AH\,Cep} \label{sec:lc:ah}

AH\,Cep is a simpler system than V453\,Cyg and V478\,Cyg because it has a 
circular orbit. However, the eclipses are shallow and provide a poor constraint 
on the ratio of the radii of the stars. There is also a light-time effect due 
to the orbit of the distant third component, but this can trivially be accounted 
for by fitting the phase offset of the primary eclipse (as done in all light 
curve analyses presented in the current work).

Extensive light curves in the Str\"omgren $uvby$ system were presented by 
\citet{bell_ahcep}, consisting of 685 observations in each filter taken 
simultaneously with a Danish photoelectric photometer on the 0.75\,m telescope 
at Observatorio del Sierra Nevada, Spain. These have complete coverage of both 
eclipses and partial coverage of the out-of-eclipse phases. Whilst having a low 
internal scatter, they suffer from systematic night-to-night offsets which are 
clearly visible in $u$ and $v$ but much weaker in $b$ and $y$. Observations are 
also available from the {\it Hipparcos} satellite \citep{perryman_hipparcos}, 
\citet{huffer_eggen} and \citet{nekrasova1960} but are significantly lower in 
quality. We therefore base our analysis on the $uvby$ light curves of 
\citet{bell_ahcep}, but also include the {\it Hipparcos} data in 
Fig.\,\ref{fig:lc:ah}. Our analysis differs from those of \cite{bell_ahcep} 
and \cite{horst_ahcep} in that we conduct a more extensive error analysis 
via the testing of multiple alternative models.

Preliminary analyses of the $uvby$ light curves indicated that the radii of 
the stars were strongly anti-correlated, a typical occurrence for shallow partial
 eclipses. This is exacerbated by the need to include third light, something 
which we know exists because a distant third star has been detected via its 
orbital motion. We therefore used our spectra to obtain a constraint on the 
light ratio of the two eclipsing stars of $0.715 \pm 0.020$, where the cooler 
and less massive star is the fainter of the two. The uncertainty in this 
measurement was determined using MCMC simulations. Our spectroscopic light 
ratio agrees well with the previous determination of $0.75 \pm 0.03$ by 
\citet{petrie1948}.

We also were able to place an upper limit on the amount of third light of 
2\% of the total light of the system from non-detection of a third set of 
absorption lines in our spectra. The errorbar is conservative. This constraint 
is suitable only for objects showing roughly similar spectral lines to the 
eclipsing stars, but alternative scenarios can be ruled out on astrophysical 
grounds. For example, if the third component was a white dwarf it would be 
extremely faint (e.g.\ a white dwarf with a \Teff\ of 100,000\,K would 
contribute approximately 0.1\% of the light of the system) and could also 
not have formed in situ because the eclipsing stars have a much shorter 
evolutionary timescale than even the most massive star capable of becoming 
a white dwarf.

Our attempts to fit for the rotational velocities of the stars failed due 
to solution indeterminacy, so we quantified the uncertainty contributed by 
stellar rotation by running solutions differing from synchronous rotation 
by 5\%. Similar difficulties were found when attempting to fit for LD 
coefficients or albedo. Our final results correspond to the best-fitting 
solution in the $b$ band for the measured spectroscopic light ratio and no 
third light (Table\,\ref{tab:wd}). The best fits are shown in 
Fig.\,\ref{fig:lc:ah}. The uncertainties in $r_1$, $r_2$ and $i$ include 
contributions due to the stellar rotation rates, albedo, LD, mass ratio, 
the uncertainty in the spectroscopic light ratio, the upper limit on the 
amount of third light, variation between solutions of the four light curves, 
and the numerical precision of the {\sc wd2004} code. All contributions were 
assessed individually then added in quadrature for each of the three parameters 
of interest.

\subsection{V453\,Cyg}

This system shows apsidal motion so the argument of periastron was included as
 a fitted parameter in all solutions. The orbital eccentricity was set to the 
value of 0.022 found from analysis of the apsidal motion of the system by 
\cite{jkt_v453}. We also fixed the mass ratio at the spectroscopic value of 
0.795 and the \Teff s at the values found in Section~\ref{sec:atmpar}.

Light curves are available in the $UBV$ bands from \cite{cohen}, totalling 
531, 532 and 512 datapoints respectively. We also obtained the $UBVR$ data 
from \citet{zakirov_v453}, containing 496 observations each, but find them 
to have a scatter higher by a factor of two, as well as a possible 
transcription problem around the egress of secondary eclipse. Our results 
are therefore based on the data from \cite{cohen}.

The total eclipses mean that these light curves are rich in information, 
especially in the relative light contributions of the two stars in the 
passbands used for the observations. As with V478\,Cyg, we found that the 
rotational velocities of the stars have a significant effect on the results, 
so we allowed for a deviation of 5\% from the measured spectroscopic value 
when determining this contribution to the overall uncertainties. We find that 
there is a strong anticorrelation between the stellar albedos and third light, 
as they are both strongly related to the amplitude of the out-of-eclipse 
variability. The amount of third light is also correlated with the LD 
coefficients, so for our final result we fixed the LD coefficients at the 
tabulated values. The amount of third light varies between fits with different 
solution options calculated with {\sc wd2004}, so we are not able to definitively 
claim its existence. Instead we suggest that it probably exists, with an upper 
limit of 7\% on its value, and that better light curves are needed to investigate 
this further.

The fixed and measured properties from the best fit of the $UBV$ light curves 
are given in Table\,\ref{tab:wd} and the best fits are shown in 
Fig.\,\ref{fig:lc:v453}. Uncertainties were assessed for the orbital inclination 
and the fractional radii of the two stars, and were the quadrature addition of 
separate contributions arising from the adopted mass ratio, treatment of LD, 
treatment of the reflection effect, numerical precision of the {\sc wd2004} code,
 and the stellar rotation rates and albedos.

\subsection{Physical properties}

Armed with the results of both spectroscopic and photometric analyses, we have 
determined the physical properties of V478\,Cyg, AH\,Cep and V453\,Cyg. For this 
we used the velocity amplitudes and orbital eccentricities from 
Table\,\ref{tab:orbit}, and the orbital inclinations and fractional stellar 
radii from Table\,\ref{tab:wd}. The calculations were performed using the 
{\sc jktabsdim} code \citep{jktabsdim} which propagates all uncertainties 
individually then adds them in quadrature to reach a final uncertainty value 
for each parameter. We note that the uncertainty in the masses of the components 
of AH\,Cep is dominated by the uncertainty in the orbital inclination, so 
obtaining better light curves of this system would also improve the mass 
measurement.

%%%%%%%%%%%%%%%%%%%%%%%%%%%%%%%%%%%%%%%%%%%%%%%%%%%%%%%%%%%%%%%%%%%%%%

\section{Discussion}

\begin{figure}
\centering
\includegraphics[width=68mm]{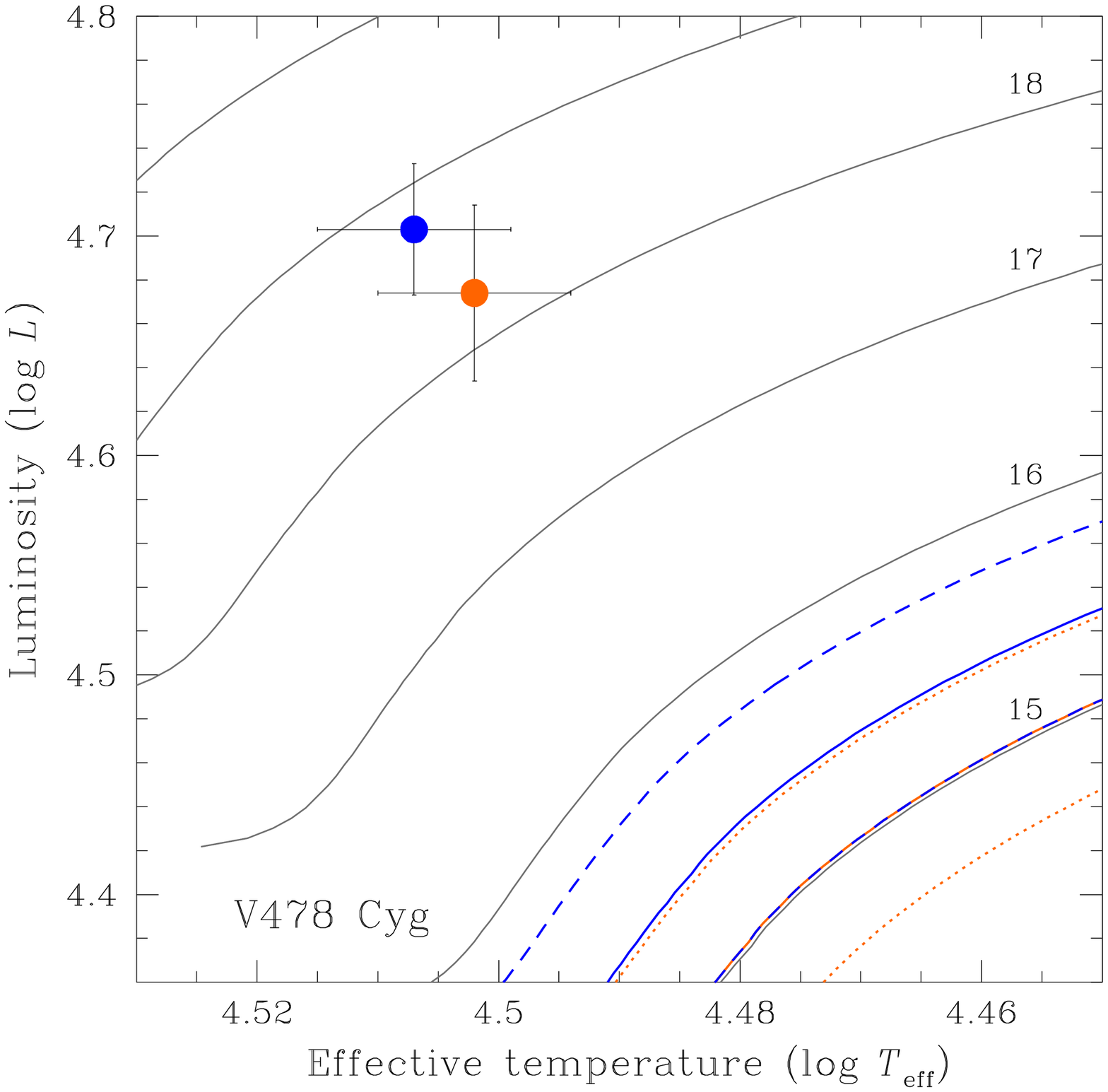} \\
\includegraphics[width=68mm]{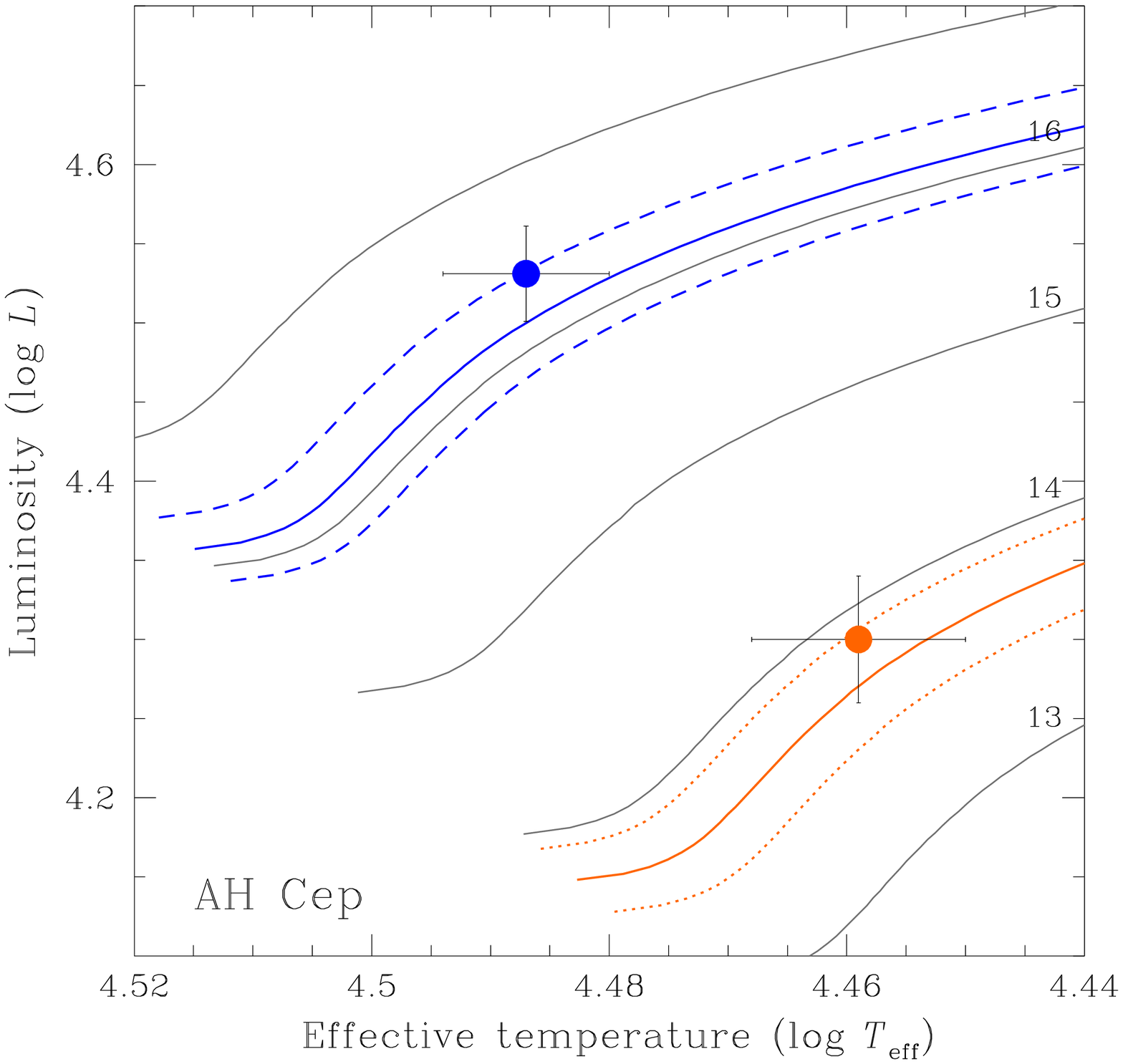} \\
\includegraphics[width=68mm]{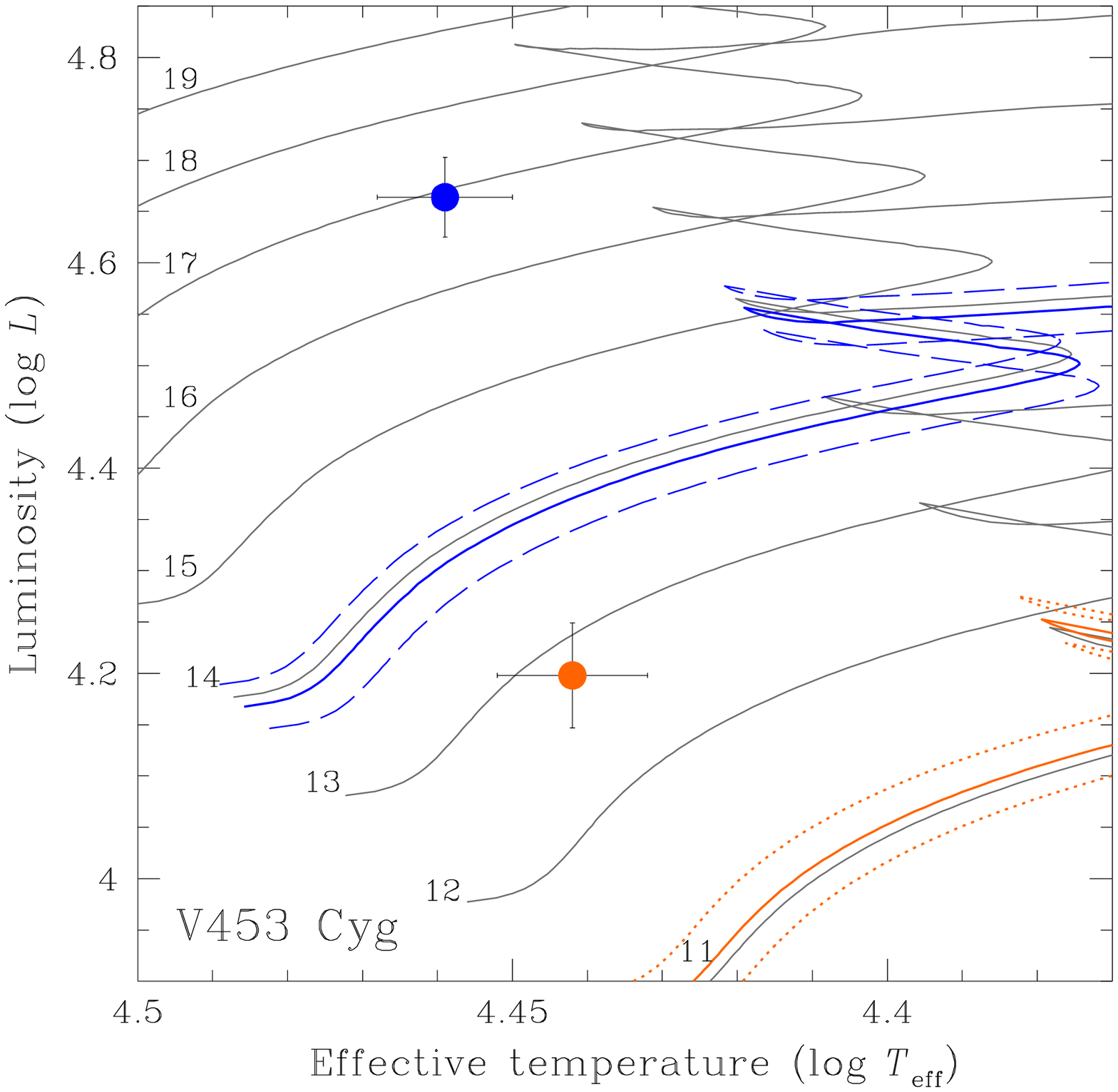} \\ 
\caption{\label{fig:plothrd}
 Positions in the HR diagram of the components for binary systems studied; 
the primary components are shown with solid circles in blue (dark grey in b\&w
print), whilst the secondaries are represented with solid circles in orange (light grey).
Evolutionary tracks are from the Geneva models (solid grey lines). Tracks for the determined
dynamical mass and assigned uncertainties are shown with solid and dashed blue (dark grey)
lines for the primary, and solid and dotted orange (light grey) lines for the secondary, 
respectively. 
All stars studied are overluminous for their dynamical
mass, or alternatively the evolutionary and dynamical masses are discrepant
 to some degree except for the components of AH~Cep which agree at
1-$\sigma$ uncetainty.}   
\end{figure}

The results of the analyses described above are summarised in
Tables \ref{tab:abs_para} and \ref{tab:abund}. We have achieved a high accuracy
in the fundamental stellar properties, with uncertainties in the masses of
1.6--2.5\% and radii of 1.0--1.9\%, giving \logg\ values to 0.009--0.021\,dex.
Having a precise \logg\ allows us to avoid the degeneracy between this quantity
and \Teff, which is estimated from the Balmer lines and He line ionisation balance,
 resulting in uncertainties of 1.7--2.5\% in \Teff. Since \Teff\ and \logg\
are the principal quantities for specifying a model atmosphere, precise values
are also very helpful in measuring chemical abundances to a high precision.
We now discuss the implications of our results for two subjects: evolutionary
models for high-mass stars, and chemical evolution in high-mass binaries.

\subsection{Evolutionary models for high-mass stars}

The accuracy achieved in the measured stellar properties of our sample is
 high enough for a detailed comparison to the predictions of evolutionary models
for high-mass stars. Moreover, determination of the photospheric abundances
is a sensitive probe on internal mixing processes. Important insights on the
internal structure of these stars could be gained from the analysis of apsidal
motion since it was observed for three of these systems (Table\,\ref{tab:sample}).
Additional information is available from their membership of stellar clusters.

Evolutionary models for high mass stars can still have difficulty in matching
observational quantities in detail. The most important stellar quantity which
governs the path of stellar evolution is mass, which can be measured in
a model-independent way in binary systems. Fig.\,\ref{fig:plothrd} shows
HR diagrams with the positions of the components of V478\,Cyg, AH\,Cep and
V453\,Cyg relative to the predictions of the Geneva models \citep{ekstrom_models};
V578\,Mon is discussed by \citet{Garcia_2014}.  The components are shown 
with filled circles, blue (dark grey in b\&w print) for the primary, and orange
(light grey) for the secondary, respectively. The theoretical evolutionary tracks 
are plotted in grey.
The evolutionary tracks for the determined dynamical masses 
are shown as solid blue/orange 
lines, with the effect of the uncertainty in mass shown with dashed/dotted blue/orange
 lines, for the primary/secondary components.

It is clear that all six stars are overluminous for their mass, although the
discrepancies for the components of AH\,Cep are within the 1$\sigma$
uncertainties. This effect was
first found for single stars by \citet{herrero}, and was termed `mass discrepancy'.
\citet{herrero} determined \emph{spectrosopic} masses from the surface gravities
and radii of stars measured from their spectra, and found them to be significantly
lower than \emph{evolutionary} masses determined by fitting evolutionary tracks
to the positions of the stars in the HR diagram.

Two of the best-documented binaries with evidence of a mass discrepancy are
V380\,Cyg \citep{guinan_v380, pav_v380, andrew_v380} and V578\,Mon
\citep{hensberge_2000,Garcia_2014}. In trying to match the observations
\citet{guinan_v380} tuned the overshooting parameter in evolutionary models,
and found a satisfactory fit for an extremely large overshooting. Once the
rotational evolutionary tracks became available for the Geneva models
\citep{ekstrom_2008} a new attempt was made to match positions of the stars
in the V380\,Cyg system \citep{pav_v380}. Rotation has almost the same effect
on stellar  evolutionary tracks as overshooting, making star more luminous, and extending
its MS lifetime. No rotational models were found that would fit the observed
properties satisfactorily. \citet{pav_v380} also determined the photospheric
abundances for the primary component, which did not show any peculiarities
in terms of the changes in abundances of the CNO elements, as expected for
rotationally-induced mixing in high mass stars (see below). This system was
revisited by \citet{andrew_v380}, who analysed the {\it Kepler} satellite
light curve plus 420 high-S/N \'echelle spectra. A grid of evolutionary models
parametrised with the overshooting parameter and an initial rotational
velocities were calculated with the {\sc mesa} code \citep{mesa}. Comparison
of the observed properties of the binary components to a new grid of evolutionary
models did not settle the mass discrepancy, showing that the theoretical models
are still inadequate and lack a significant amount of near-core mixing.

The mass discrepancy for high-mass stars in binary systems is now
a well-determined effect, which points to shortcomings in modern evolutionary
models. The inclusion of some extra mixing into the models allows to minimise
the discrepancy between the dynamical and evolutionary masses of stars but,
as stressed by \citet{andrew_proc}, the high overshooting values appear
to be inconsistent with asteroseismological values known for single stars
of similar masses.
 As already stressed in Sect.\,1 the calculations by \citet{pedersen}
have shown to be very promising for probing efficiency and profile of internal mixing
from constraints between gravity-mode oscillations and surface abundances,
in particular nitrogen contents. 
 We postpone further discussion of this important issue
for a dedicated study which will involve a sample of 12 or more well-studied
binary systems (Tkachenko et al., in prep.).

\subsection{Chemical evolution in high-mass binaries}

The principal energy source in high-mass stars is the CNO cycle. In the beginning 
of the process, N is produced in the CN cycle at the expense of C. Later, when 
the CNO cycle reaches equilibrium, additional N is produced, at the expense of 
both C and O. Thus N enhancement, C depletion, or preferably the N/C abundance 
ratio have been recognised as sensitive indicators of the evolutionary processes 
in high-mass stars \citep{przybilla_cno,maeder_cno}. The photospheric chemical 
composition, and in particular the N overabundance, becomes a sensitive probe 
of mixing processes in stellar interiors. Rotation has been postulated as an 
efficient mechanism for enhanced mixing and transport of a processed material, 
with N and He enhancement, up to the stellar surface layers. Stellar evolutionary
 models with rotationally induced mixing predict a causal relationship between 
initial stellar rotational velocities and the N/C abundance ratio \citep[][and 
references therein]{maeder_araa,langer_araa}.

In a seminal work on the CNO abundances for a large sample of Galactic OB stars 
\citet{gies_cno} revealed a variety of abundance peculiarities, including N 
enhancement. In a dedicated large-scale observational project using multi-object 
spectroscopy, OB stars in eight stellar clusters in the Galaxy, LMC and SMC, 
were observed \citep{evans_flames1,evans_flames2}. Studies of the N abundances 
for OB stars in the Galaxy and the Magellanic Clouds have become controversial 
since no simple and clear relationship between the N abundance and projected
rotational velocity was disclosed \citep{hunter_letter,hunter_paper}. Instead three 
distinctive groups of stars have been revealed (to be discussed at the end of this 
section).

\begin{figure*}
\centering
\includegraphics[width=58mm]{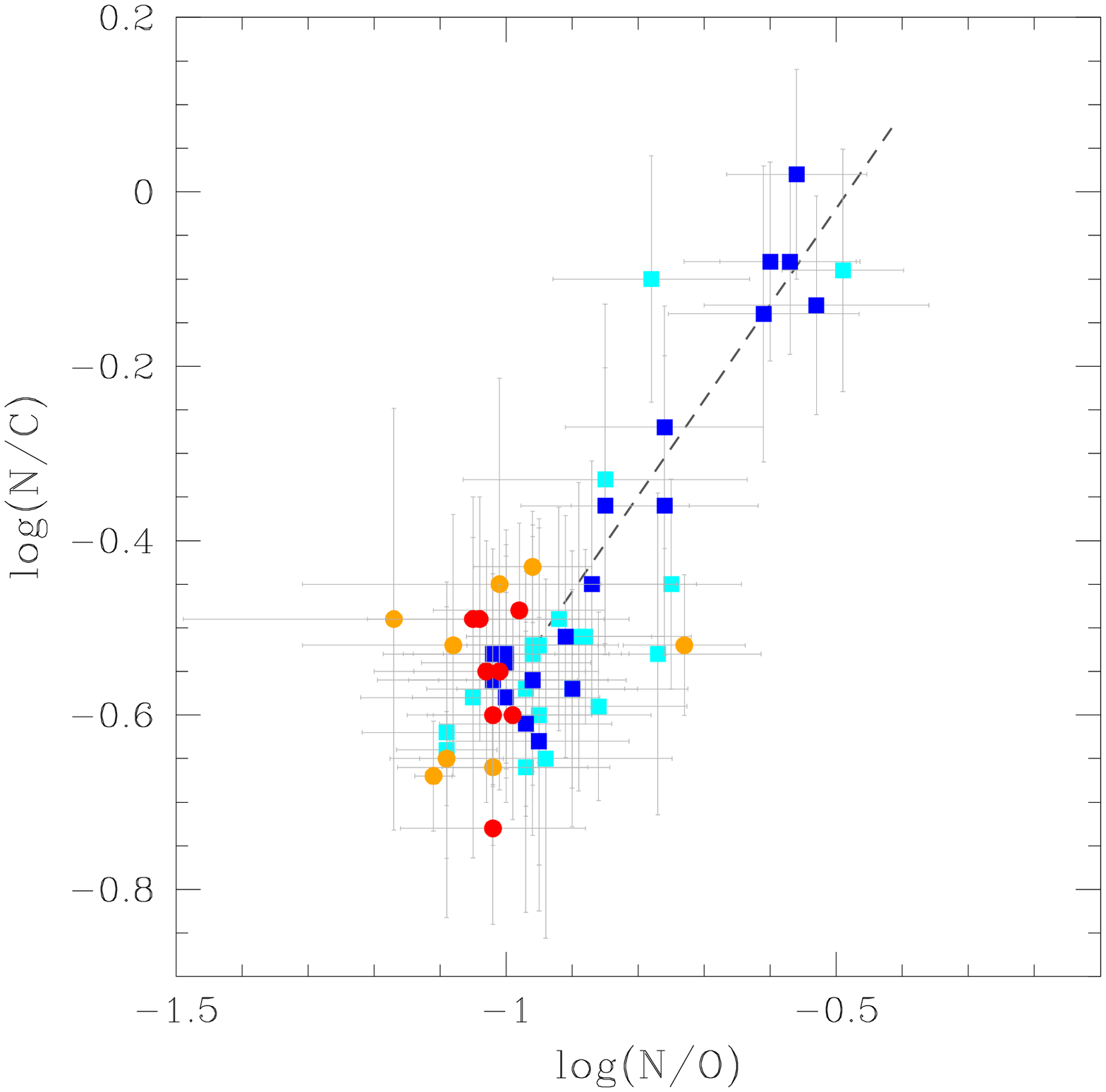} 
\includegraphics[width=58mm]{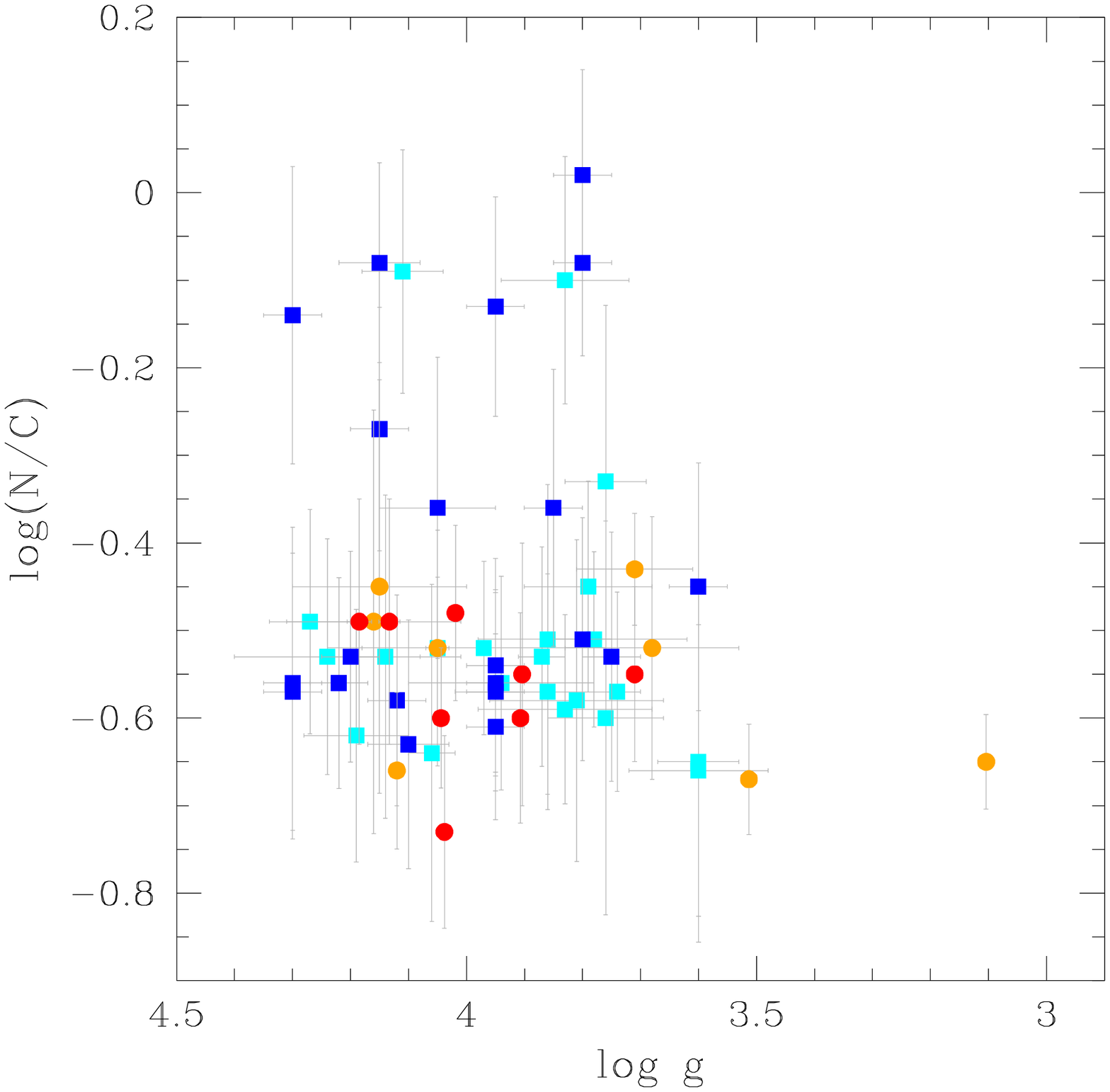}  
\includegraphics[width=58mm]{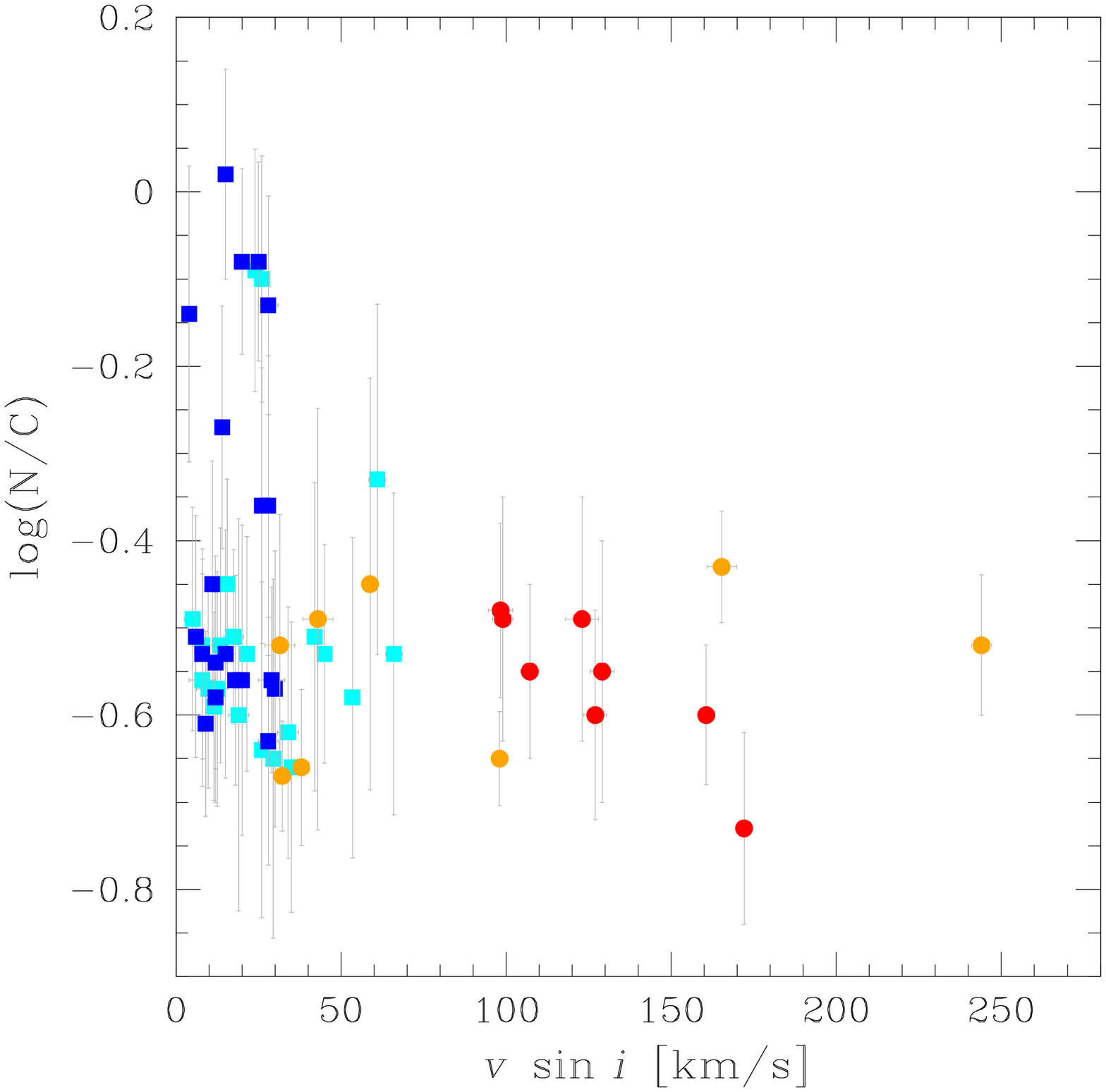} 
\caption{\label{fig:cno}  Left panel: abundances of the CNO elements for high-mass
 stars shown in the logaritmic (N/C) vs (N/O) diagram. The components in binary systems 
are represented by solid circles, in red (dark grey in b\&w print) from our sample, and in
orange (light grey) from previous studies. The results of  
abundance determinations for single early-B type stars are represented by solid squares,
in blue (dark grey) from  \citet{fernanda_standard}, and in
cyan (light grey) from \citet{lyubimkov_cno}. Single stars obey a trend indicated by analytical
approximation to the nuclear reactions path for CN-cycle derived in \citet{przybilla_cno} and
\citet{maeder_cno}. It is evident that their relatives in binary systems occupay narrow region around
the initial values with log(N/C) $\simeq -0.54$ and log(N/O) $\simeq -0.97$ \citep{fernanda_standard}.  
Middle panel: evolutionary changes in log(N/C) versus \logg. The meanings of the symbols
 are the same as in the left panel. Apparently,  evolution does not make any change in the surface
abundance of CNO elements. 
Right panel: Dependence of nitrogen-to-carbon ration on the projected
rotational velocity  (the meanings of the symbols are the same as in
the left panel). The samples for single early-B type stars are biased toward sharp-lined stars.
No one of 16 binary components is showing pronounced  N/C even those with high rotational 
velocities.}
\end{figure*}

Determination of the elemental abundances for fast-rotating stars is 
observationally challenging. Spectral lines overlap, and fast rotation broadens 
and weakens them. Studies of slowly rotating early-B type stars allowed 
\citet{przybilla_cno} and \citet{fernanda_standard} to achieve much more accurate 
line profile measurements 
and also more accurate atmospheric parameter measurements through the ionisation 
balance for several elements. They found a tight correlation of the abundance 
ratios N/C versus N/O. Their finding was further strengthened in a subsequent 
analysis by \citet{maeder_cno}, who re-examined part of an extensive 
{\sc vlt/flames} sample. This correlation closely follows theoretical 
predictions due to nuclear processing in the CN branch of the CNO cycle 
\citep{przybilla_cno, maeder_cno}. 
{ The enhancement of the
 nitrogen in high-mass stars is consequence of a bottle-neck nuclear reaction
chain in the CN cycle, and it happens on the expense of the carbon.  On these theoretical
grounds \citet{przybilla_cno} and \citet{maeder_cno} predicted the slope d(N/C)/d(N/O) = 3.77.
This analytical approximation to the nuclear reactions path beginning at initial
 initial ratios N/C $\simeq 0.31$ and N/O $\simeq 0.11$ is represented by dashed line in
the left panel of Fig.\,\ref{fig:cno}. It is obvious that abundance ratios of the 
CNO elements for single
early-B type stars from samples of \citet{fernanda_standard} 
(solid squares in blue) and \citet{lyubimkov_cno} (solid squares in cyan) obey
theoretically predicted evolutionary changes. However, the mechanism which brings 
nuclear-processed material to the stellar surface layers becomes more obscure 
because the majority of the stars examined are slow rotators (see the right panel in
Fig.\,\ref{fig:cno}).

\citet{proffitt_boron} investigated rotationally driven mixing in high mass 
stars through another channel. Since boron is destoyed at much lower temperatures 
than needed for the CN cycle, it has diagnostic power \citep{fliegner, 
frischknecht}. HST spectroscopy of \ion{B}{iii} lines in the UV spectral range 
showed boron depletion in a sample of early-B stars in the young open cluster 
NGC\,3293, but with the efficiency at or slightly below the low end of the range
 predicted by rotating evolutionary models \citep{proffitt_boron}.

An important study towards the identification of possible mechanisms which 
enable an early change in the photospheric N abundance in high-mass stars 
was undertaken by \citet{conny_nitro}. They collected a statistically 
significant sample of well-studied Galactic B-stars for which seven 
observables were available (surface N abundance, rotational frequency, 
magnetic field strength, and the amplitude and frequency of their dominant 
acoustic and gravity modes of oscillation). A multivariate analysis indicated 
that the \Teff\ and the frequency of the dominant acoustic oscillation mode 
have the most predictive power of the surface N abundance.

How do the results of our photospheric CNO abundance measurements fit in the 
picture outlined for single stars?  An examination of the abundances for the 
eight high-mass stars in dEBs studied in the present work (Table\,\ref{tab:abund})
 shows a little dispersion clustered around the initial values established for 
high mass stars (see Fig.\,\ref{fig:cno}, left panel, solid cyrcles in red). 
Since the dispersion of the derived 
abundances is small, we have calculated the mean values, also given in 
Table\,\ref{tab:abund}. Comparison to the `present-day cosmic' abundances 
\citep{fernanda_standard} shows agreement within 1$\sigma$. Our sample of 
eight high-mass stars spans a large parameter space with the masses between 
10 and 16 \Msun, \Teff s between 25\,000 and 32\,000\,K, \logg s from 3.7 
to 4.2\,dex, and $v\sin i$ values from 100 to 180\kms. In spite of a broad 
range of stellar quantities, no substantial changes which would follow expected 
theoretical predictions are found 
 (the left panel of 
Fig.\,\ref{fig:cno}). This is in a plain contrast to the results found  from 
the most accurate determination of CNO abundances for sinle early-B type stars
by \citet{fernanda_standard} (solid blue squares in the
Fig.\,\ref{fig:cno}), and corroborated by \citet{lyubimkov_cno}
(solid cyan squres in Fig.\,\ref{fig:cno}). It should be noted that samples
for single stars 
represented in Fig.\,\ref{fig:cno} contain stars covering a similar parameter
space. In sample of \citet{fernanda_standard} the masses
(derived spectroscopically) are between 7 and 19 \Msun, \Teff s between 17\,000 
and 33\,000\,K, \logg s from 3.6 to 4.3\,dex, and $v\sin i$ values from 4 to 30 \kms.    
\citet{lyubimkov_cno} sample comprises stars at a low end of high mass domain, and
an upper part of the intermediate mass stars. The masses  are between 5.5 and 11.2 \Msun, 
\Teff s between 15\,400 
and 24\,000\,K, \logg s from 3.6 to 4.3\,dex, and $v\sin i$ values from 5 to 66 \kms.
Both samples are biased toward low projected rotatioal velocities, more strongly
for the stars in \citet{fernanda_standard} than \citet{lyubimkov_cno},
for achieving a high accuracy in determination of the atmospheric parameters and
abundances. }

 The majority of the stars in 
our sample are young: only V453\,Cyg\,A is beyond the halfway point of its MS 
lifetime. The middle panel of Fig.\,\ref{fig:cno} shows the N/C abundance ratio as a function of 
\logg, the latter being a good indicator of stellar evolution. In this 
representation an absence of evolutionary changes of the N abundance is even 
more evident, in particular compared to the sample of single early B-type stars. 
Also, it is 
striking that N enhancement is present for single high mass-stars early in 
their MS lifetime.

Fig.\,\ref{fig:cno} includes other high-mass stars in binaries (solid circles in orange): 
V380\,Cyg 
\citep{andrew_v380}, $\sigma$\,Sco \citep{andrew_sigma}, $\alpha$\,Vir 
\citep{andrew_spica}, HD\,165246 \citep{mayer_hd}, and V621\,Per 
(Southworth et al., in prep). Of particular interest are V380\,Cyg\,A,
 and V621\,Per\,A, which are evolved stars, with \logg s of 3.1 and 3.5\,dex, 
respectively. Despite their advance evolutionary stage, these evolved stars 
do not (yet) show any photospheric abundance anomalies in terms of N enhancement 
or C depletion. $\alpha$\,Vir\,A and $\sigma$\,Sco\,A are $\beta$\,Cep-type 
pulsators, and are in an advanced stage of their MS lifetimes without any 
trace of changes in the CNO abundances. HD\,165246\,A is the most massive 
star in this extended sample, with $M = 20$\Msun, and with the highest 
$v \sin i$ of 230\kms\ \citep{mayer_hd}. Recently, its oscillating nature 
was found by \citet{johnston} using {\it Kepler} data. No outlier is seen 
among this group of high-mass stars, which corroborates our findings in the 
current work.  It should be noted that the CNO abundances for these binaries 
are determined with the same methodology, and NLTE spectrum analysis, 
as performed in this work.

 If rotational mixing is the main physical mechanism which brings the CNO nuclearly
processed material to the stellar surface layers then the changes in the nitrogen-to-carbon
ratio should grow with  stellar rotational velocity. The right panel in Fig.\,\ref{fig:cno}
shows dependence of log(N/C) on the projected rotational velocity $v \sin i$ for single stars (squares)
and binary components (circles). As noted early, the samples for single stars are
biased to low $v \sin i$ for the stars in \citet{fernanda_standard}, and up to moderate
 $v \sin i$ in \citet{lyubimkov_cno}. Our sample is not biased to $v \sin i$ and covers
binary components with $v \sin i$ from 100 to 170 \kms (this work), and from 32 to 244 \kms
(compilation of the previous studies).
 Plot (the right panel in Fig.\,\ref{fig:cno}) is reminiscent of the diagram first
presented in \citet{hunter_letter, hunter_paper}.
Examing theoretically predicted dependence of the nitrogen abundance on the projected
rotational velocity for OB stars, both in the Galaxy and the Magellanic Clouds, they
 disclosed three distinct
 groups: (i) stars with a low $v \sin i$ with high N abundance, (ii) fast rotators with no
N enhancement, and (iii) stars which show correlation between N enrichment and the projected
 rotational velocity. Evidently, single B-type stars in the right panel of Fig.\,\ref{fig:cno}
show a large  spread in the N abundance and resemble group (i) in the quoted diagram
of \citet{hunter_letter, hunter_paper}.  Next, the stars in binaires, with a large
spread in the rotational velocities (with the inclination angle known from the light curve
analysis for the stars in binaries a true rotational velocity could be calculated) occupied
domain of group (ii). Not present in our plot  are stars, either singles or binary
components, from group (iii). The N enriched early-B type stars with high $v \sin i$
are either missing for intrinsic reason(s) in binary systems, or they are missing
for single stars because of an observational bias. Still, it is not excluded some  stars
with low $v \sin i$ are highly rotating star seen close to pole-on. 
%However,
%caution is needed because we are dealing with a small number statistics.
 Recently progress in the measurements of CNO abundance for highly rotating stars
(high $v \sin i$) was made  \citep{cazorla2017a, cazorla2017b, markova}.
But these new observational studies are dealing with stars in the mass range
$M > 20$ \Msun, thus higher than we are considering  in the present work, hence both
samples can not be directly compared. Still, it is worth mentioning that these new
observational studies did not support rotational mixing as only process governing the
changes in the surface chemical composition in high-mass single  stars.

\section{Summary and Conclusions}

Our understanding of the structure and evolution of high-mass stars is still 
incomplete, despite intensive theoretical and observational effort. The aim of 
our project is to extend the number of OB binaries with precise and accurate 
measurements of their physical properties, to become a statistically significant 
sample able to provide an advance in probing the predictions of theoretical 
models.

We presented and analysed new high-S/N \'echelle spectroscopy for three OB 
binary systems: V478\,Cyg, AH\,Cep and V453\,Cyg. Using the method of spectral 
disentangling we obtained the spectroscopic orbits and individual spectra of 
the component stars. The spectroscopic orbits were combined with a detailed 
analysis of published light curves in order to determine the masses and radii 
of the stars. Their disentangled spectra were analysed to determine the 
atmospheric parameters and photospheric chemical abundances. The uncertainties 
in the measured masses are lower than 2.5\%, in the radii below 1.9\%, and in 
the \Teff s below 2.5\%.

Accurate stellar data allows a robust testing of evolutionary models for 
high-mass stars. We find that no star in our sample matches the evolutionary 
track for its dynamically-measured mass. All stars studied have a higher 
luminosity than predicted for their dynamical masses, showing that the 
`mass discrepancy' identified by \citet{herrero} still exists. The deviations 
between the dynamical and evolutionary masses for stars in our sample range 
from 1\% to 10\%. Thus results corroborate the mass discrepancy found for 
single stars as well for OB binaries (see \citealt{andrew_proc} and 
references therein). The source(s) of the mass discrepancy remains hidden 
and absent from theoretical models. Further investigation of this problem 
will be found in a forthcoming study (Tkachenko et al., in preparation).

An important aspect of our work is determination of the photospheric chemical 
abundances of  C, N, O, Mg and Si. We also included the dEB V578\,Mon 
\citep{Garcia_2014} in this analysis, superseding our previous analysis 
based on more limited spectroscopic data (PH05). Similarly, our new results 
for V453\,Cyg supersede those presented by \citet{pav_v453}, which were based 
on spectra of much smaller wavelength coverage.

Whilst global stellar quantities for single and binary stars share the same 
tendency of overluminosity or mass discrepancy, this appears not to be the 
case for the surface CNO abundance pattern. The CNO abundances of stars in 
our sample cluster around $\log ($N/C$) = -0.56 \pm 0.06$ and $\log ($N/O$) 
= -1.01 \pm 0.06$ (Table\,{\ref{tab:abund}), whilst the `present-day cosmic  
standard' values
determined by \citet{fernanda_standard} from a sample of 
sharp-lined early-B stars are: $\log ($N/C$) = -0.54 \pm 0.06$ and $\log ($N/O$) 
= -0.97 \pm 0.06$. The  initial values for single B-stars, 
and for OB stars in binary 
systems are the same within their uncertainties. High-mass stars in binaries 
do not show any changes in the course of their evolution, or as a function of 
stellar properties (mass, \logg, $v\sin i$). This result corroborates our 
previous abundance determinations for several OB binaries (PH05, 
\citealt{pav_v453, pav_v380, mayer_hd, andrew_v380, andrew_sigma, andrew_spica}).
 However, caution is needed since all these stars are at the low-mass end of 
the high-mass category, with masses 10--16\Msun. We note that \citet{mayer_hd} 
found no abundance anomaly for HD\,165234\,A, which has a mass of $25 \pm 3$\Msun.
 Since the recent discovery of its pulsations \citep{johnston}, this system is 
promising target for an improvement in the previous analysis, due to the synergy 
with asteroseismology. Recently, \citet{martins_mahy} determined the CNO 
abundances for the binary systems DH\,Cep (38 and 33\Msun) and Y\,Cyg (16 
and 17\Msun). The results for Y\,Cyg fall around the values we found for our 
sample. In the case of DH\,Cep, abundances are available for the secondary 
component only. These are $\log ($N/C$) = -0.21 \pm 0.55$ and $\log ($N/O$) 
= -1.11 \pm 0.25$. The prohibitively large errors unfortunately preclude 
further discussion of these results.

Usually, it is taken for granted that evolution for the components in detached 
binaries is not affected by the presence of the companion, so is the same as 
for single stars, until proximity effects become important once the stars have 
evolved to larger radii. In the present work, we arrived at two apparently 
contradicitory findings: (1) the binary components also show the overluminosity 
found for single stars, usually referred to as the mass discrepancy; and (2) 
the photospheric chemical composition appears unaffected by evolution on the 
MS, which is different to single stars. Whilst a source of an extra luminosity 
for high-mass stars still should be identified and correctly accounted for in 
theoretical models, probably as some mechanism for causing extra mixing in 
stellar interiors, the second finding we reported in this work points to some 
(efficient) mechanism of blocking or damping diffusion of nuclear-processed 
material to the stellar surface. It is not surprising that the structure and 
evolution of stars in binaries is affected by tidal effects due to the proximity 
of a companion. Important issue is the fact that fundamental stellar quantities 
are calibrated with detached binaries. Therefore, further observational 
clarification of these open issues are needed, in particular with an extension 
to a higher mass range than covered in this work.

%%%%%%%%%%%%%%%%%%%%%%%%%%%%%%%%%%%%%%%%%%%%%%%%%%%%%%%%%%%%%%%%%%%%%%

\section*{Acknowledgements}

We thank the anonymous referee for the timely and constructive comments
that have helped to improve the paper.
Also, we are grateful to Kelsey Clubb for obtaining and reducing the spectra of 
V478\,Cyg, and to Laurits Leedj\"{a}rv and Mamnun Zakirov for communicating 
their photometric obsrvations of binaries studied in this work. KP and ET are 
financially supported by the Croatian Science Foundation through grant 
IP 2014-09-8656. We acknowledge help in the course of this work from 
A.~Dervi\c{s}o\u{g}lu, V.~Kolbas and  K.~Tisani\'{c}.

Based on observations collected at the Centro Astron\'omico Hispano Alem\'an 
(CAHA) at Calar Alto, operated jointly by the Max-Planck Institut f\"ur 
Astronomie and the Instituto de Astrof\'{\i}sica de Andaluc\'{\i}a (CSIC), 
and on observations made with the Nordic Optical Telescope, operated on the 
island of La Palma jointly by Denmark, Finland, Iceland, Norway, and Sweden, 
in the Spanish Observatorio del Roque de los Muchachos of the Instituto de 
Astrof\'{\i}sica de Canarias, and Lick Observatory, USA.

%\expandafter\ifx\csname natexlab\endcsname\relax\def\natexlab#1{#1}\fi

%%%%%%%%%%%%%%%%%%%%% REFERENCES %%%%%%%%%%%%%%%%%%
\bibliographystyle{mnras}
% \bibliography{references}
% Don't change these lines
% \bsp    % typesetting comment

\label{lastpage}

\end{document}